\documentclass[]{JHEP3}
\usepackage{graphicx}
\usepackage{amsmath}
\numberwithin{equation}{section}
\preprint{Cavendish--HEP--11/05}
\title{NLO QCD Corrections to Graviton Induced Deep Inelastic Scattering}
\author{W.J.~Stirling, E.~Vryonidou\\Cavendish Laboratory, J.J. Thomson Avenue, Cambridge CB3 0HE, UK}

\abstract
{We consider Next-to-Leading-Order QCD corrections to ADD graviton exchange relevant for Deep Inelastic Scattering experiments. We calculate the relevant NLO structure functions by 
calculating the virtual and real corrections for a set of graviton interaction diagrams, demonstrating the expected cancellation of the UV and IR divergences. We compare the NLO and LO results at the centre-of-mass energy relevant to HERA experiments as well as for the proposed higher energy lepton-proton collider, LHeC, which has a higher fundamental scale reach.}
\begin{document}

\tableofcontents
\section{Introduction}
Models with extra dimensions have been proposed to address the hierarchy problem. These can in general be classified into two groups depending on the geometry of the extra dimensions; flat or factorisable models and warped or non-factorisable models. The former includes the model proposed in \cite{ArkaniHamed:1998rs,Antoniadis:1998ig} by Arkani-Hamed, Dimopoulos and Dvali (ADD) and its variants, while the latter includes the Randall--Sundrum (RS) model \cite{Randall:1999ee}. 

The large extra dimensions model introduced in \cite{ArkaniHamed:1998rs,Antoniadis:1998ig} suggests that the Standard Model (SM) particles live in the usual 3+1-dimensional space, while gravity propagates in a higher D-dimensional space. The weakness of gravity with respect to other forces is imposed by the large size of the compactified extra dimensions. In this model Newton's constant is expressed  as
\begin{equation}
G_N^{-1}=M_{P}^2=8\pi R^{n} M_{D}^{2+n},
\label{Planck}
\end{equation}
where $M_P$ is the Planck mass, $M_D\sim$TeV is the fundamental mass scale, $n$ is the number of extra dimensions and $R$ the radius of the compactified space, assumed to be a torus in this model. The size of the extra dimensions is determined by their number, $n$, and the fundamental mass scale, $M_D$, through Eq.~\ref{Planck}. For a fundamental scale of the order of 1~TeV and one extra dimension, its size would be of the order of the solar system size, causing large deviations from the inverse square law of gravitation. Therefore one extra dimension is already ruled out by experiments. For more extra dimensions, the size gets rapidly smaller, not contradicting submillimetric gravity measurements \cite{Kapner:2006si}.
 
 In the ADD model \cite{ArkaniHamed:1998rs}, the graviton corresponds to the excitations of the D-dimensional metric. These can be expressed as a tower of Kaluza-Klein (KK) modes. The interaction Lagrangian for gravitons and the SM fields is given by
\begin{equation}
\mathcal{L}_{int}=-\frac{1}{\overline{M}_{P}}\sum_{\vec{n}}G^{\vec{n}}_{\mu\nu}T^{\mu\nu},
\label{Lag}
\end{equation}
with $\overline{M}_{P}=M_P/\sqrt{8\pi}$ the reduced four-dimensional Planck mass, $T_{\mu\nu}$ the energy momentum tensor of the SM fields and $\vec{n}=(n_1,n_2,...,n_{n})$ a $n$-dimensional vector of integers labelling the massive gravitons. A set of Feynman rules for the interactions of these massive KK states with the Standard Model particles was presented in \cite{Giudice:1998ck, Han:1998sg}. Note that this is an effective theory of interactions valid in the cisplanckian region, that is when the centre-of-mass energy of the parton collision is smaller than the fundamental scale $M_D$. The predictions of this effective theory fail as we approach the quantum gravity scale. As we reach the planckian region where $\sqrt{s}\simeq M_D$, we need theoretical input from quantum gravity, and so it is presently impossible to predict experimental signals reliably.

The KK resonances have masses equal to $m_{(\vec{n})}=|\vec{n}|/R$. This results in a small mass gap of order $R^{-1}$, e.g. for one extra dimension of size 1~$\mu$m the mass gap is $\mathcal{O}$(1~eV). This renders different masses practically indistinguishable and permits replacing the sum over discrete mass values by an integral over a continuum with a given density of states. This density of states is obtained by considering the number of modes with KK index between $|n|$ and $|n|+dn$, and is given by 
\begin{equation}
\rho(m)=\frac{dN}{dm}=S_{\delta-1}\frac{\overline{M}_{P}^2}{M_D^{2+\delta}}m^{\delta-1}.
\label{density}
\end{equation}

As seen from the Lagrangian in Eq.~\ref{Lag}, graviton interactions are suppressed by inverse powers of $\overline{M}_{P}$. Nevertheless, the summation over the large number of accessible KK modes, that is the integration over the density in Eq.~\ref{density}, cancels the dependence on $\overline{M}_P$.
As an example, the inclusive graviton production cross section, 
$\sigma_m$, is expected from the graviton couplings to be proportional to $\overline{M}_P^{-2}$, which combined with Eq.~\ref{density} exactly cancels the dependence on $\overline{M}_P$. This leads to an effective interaction suppressed by inverse powers of the fundamental mass scale $M_D$, thus giving observable effects for $M_D$ near the TeV scale.

Interesting phenomenological implications of this extra dimensions model have been studied extensively in the literature. 
These include real graviton emission and virtual graviton exchange. A set of processes, such as graviton plus gauge boson production, was studied in \cite{Giudice:1998ck,Han:1998sg}, where the Feynman rules were first presented. The experimental signature for graviton production is missing energy, as decay into SM particles is suppressed by a factor of $1/M_P^2$, which is not compensated by phase space. Therefore gravitons behave as heavy and stable particles, once produced. One should expect missing energy signals at both lepton and hadron colliders. This missing-energy signal does not correspond to a fixed invisible-particle mass as the graviton has a continuous distribution in mass. This differentiates graviton searches from other Beyond the Standard Model (BSM) physics searches, such as supersymmetry which would also give a missing energy signal. The experimental signatures depend strongly on the number of extra dimensions, $n$, and the fundamental mass scale, $M_D$,  and so we should be able to determine or at least constrain both of these model parameters. Virtual graviton effects will cause deviations from SM predictions for fermion and boson pair production. The cross sections are divergent at tree level, and thus one has to introduce an ultra-violet cut-off that is usually taken to be of the order of the fundamental scale \cite{Giudice:1998ck}. The search for virtual graviton effects is complementary to the production search, and will shed light on this cut-off and its dependence on the fundamental scale and the number of extra dimensions.

Both the LEP and Tevatron experiments have set limits on $M_D$.
The combined results of the LEP experimental collaborations for the monophotons channel are summarised in \cite{Ask:2004dv}. The CDF and D0 collaborations have searched for signals in the monojets and monophotons channels for real emission~\cite{Aaltonen:2008hh} as well as dilepton and diphoton channels for virtual exchange effects. For a brief summary of the experimental searches and the current constraints on the fundamental scale and the number of extra dimensions see \cite{Landsberg:2008ax} and references therein. A more recent study of the dijet angular distribution in D0 was performed in \cite{:2009mh}. The most stringent current constraints on the fundamental scale are set at 1-1.5~TeV, but these also depend on the number of extra dimensions. The LHC is expected to probe extra dimensions effects up to higher scales, due to the larger centre-of-mass energy.  Recently, a first study of the bounds set on the cut-off scale from the dijet cross section measured at the LHC  \cite{Collaboration:2010eza,Khachatryan:2010te} was presented in \cite{Franceschini:2011wr}, which are in the few TeV region.

Recently, several graviton processes were calculated at Next-to-Leading Order (NLO) in QCD. Leading order calculations suffer from uncertainties due to the choice of renormalisation and factorisation scales. NLO results are less dependent on the scale choice and thus improve theoretical predictions. Calculations have been performed for both ADD and RS models. Here we just mention a few of the processes studied at NLO QCD: graviton production \cite{Karg:2009xk,Li:2006yv}, graviton and photon associated production \cite{Gao:2009pn}, graviton plus two jets production \cite{Hagiwara:2008iv}, heavy resonance graviton production and decay into top quarks \cite{Gao:2010bb}, diphoton production \cite{Kumar:2008pk}, graviton plus Z production \cite{Kumar:2010kv}, Drell-Yan \cite{Mathews:2004xp} and gravitational scattering at transplanckian energies \cite{Lodone:2009qe}.

Graviton contributions to Deep Inelastic Scattering (DIS) have been considered by HERA experiments. The absence of any deviation from the SM prediction was used to set limits on the fundamental mass scale. The most recent limits for $e^+ p$ collisions from the H1 experiment are given in \cite{H1prel}, while more information about the analysis method can be found in \cite{Adloff:2003jm}. These studies use the leading-order (LO) prediction for the signal and NLO QCD corrections are not taken into account. Given the current limits on the fundamental scale and the centre-of-mass energy available at the HERA experiments, discovery of graviton effects does not fall within the experimental reach.

The graviton contribution to DIS will be more important in the context of the proposed Large Hadron electron Collider~(LHeC) \cite{Zimmermann:2010zzd}. The proposed design suggests using the LHC ring and a linear electron accelerator to collide 7~TeV protons with 140~GeV electrons. The significant increase of the centre-of-mass energy will allow much higher momentum transfers to be probed, with the effect of graviton DIS becoming important and much higher mass scales becoming accessible.

In this paper, we evaluate the real and virtual corrections for simple graviton exchange relevant for DIS experiments. We present analytic results for the relevant structure functions at LO and NLO. We then present numerical results on the differential cross section and comment on the effect of NLO QCD corrections at HERA and LHeC energies.

\section{Leading order calculation}
\subsection{Structure functions definition}
\begin{figure}
\centering
\includegraphics[scale=0.8] {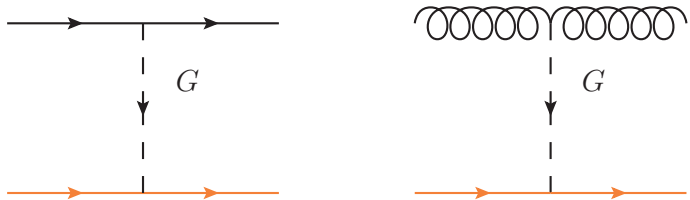}
\caption{Diagrams for $q\ell\rightarrow q\ell$ and $g\ell\rightarrow g\ell$. Red denotes the lepton line.}
\label{DISdiag}
\end{figure}

The leading order results for graviton exchange in DIS experiments were first published in \cite{Mathews:1998qn} soon after the Feynman rules for gravitons became available. The corresponding parton-level Feynman diagrams for graviton exchange are shown in Fig.~\ref{DISdiag}. We note that there is a LO gluon diagram contributing to the scattering, in contrast to photon mediated  DIS where gluon constituent scattering can only occur at NLO QCD. 

We start by considering the partonic differential cross section for the exchange of a graviton between a quark and a lepton, given by:
\begin{equation}
 \frac{d^2\hat{\sigma}}{dQ^2}=\frac{\pi}{32M^8_S}\frac{Q^4}{y^2}[32-64y+42y^2-10y^3+y^4],
\end{equation}
with $Q^2$ and $y$ the usual DIS variables and $M_S$ the cut-off introduced to regularise the UV divergence of the graviton exchange.
To obtain the hadronic cross section, the partonic cross section is convoluted with the proton parton distribution functions which in this case are sampled at momentum fraction $x$, giving the double differential cross section:
\begin{equation}
 \frac{d^2\sigma}{dQ^2dx}=\frac{\pi}{32M^8_S}\frac{Q^4}{y^2}[32-64y+42y^2-10y^3+y^4]\sum_q[q(x)+\overline{q}(x)],
\label{LOgrav}
\end{equation} where the sum is over all quark flavours, with each quark flavour contributing equally.
For comparison purposes we write down the corresponding expression for photon exchange in unpolarised deep inelastic scattering using the same variables:
\begin{equation}
 \frac{d^2\sigma}{dQ^2dx}=\frac{4\pi\alpha^2}{Q^4}[y^2-2y+2]\sum_q \frac{1}{2}e_q^2 [q(x)+\overline{q}(x)],
\end{equation}where the contribution of each flavour depends now on the electric charge squared.
The difference between the two expressions originates from the different dimension operators describing the interactions of photons and gravitons with the quarks. It is customary to re-express the cross section for photon exchange for electron DIS in terms of the structure functions $F_1$ and $F_2$, with $F_3$ introduced when we also consider the $Z$ boson exchange contribution:
 \begin{equation}
 \frac{d^2\sigma}{dQ^2dx}=\frac{4\pi\alpha^2}{xQ^4}[xy^2F_1+(1-y)F_2+y(1-\frac{1}{2}y)xF_3],
\end{equation}with the sum over parton distribution functions now within the structure functions and $F_1$, $F_2$ and $F_3$ receiving contributions from pure $\gamma$ and $Z$ exchange and their interference.

We will now proceed to find the equivalent expression for graviton mediated DIS. The advantage of this reformulation of the cross section will be exploited in the next section as it facilitates the analytic calculation of the NLO cross section. As with photon DIS the starting point is the hadronic tensor. This now has four indices and contains all the information about the interaction of the gravitational current with the proton target P with spin $S$:
\begin{equation}
 W_{\mu\nu\alpha\beta}=2\pi^2\int d^4z\, e^{iqz}\langle P,S|T_{\mu\nu}(z)T_{\alpha\beta}(0)|P,S\rangle,
\end{equation}with $T_{\mu\nu}$ the energy momentum tensor which describes the coupling of graviton to other SM particles. In analogy with the photon case, this hadronic tensor is conserved and can be decomposed in term of structure functions. The idea was first introduced in \cite{Lam:1981tg}, where the authors consider a DIS thought-experiment in which the protons are bombarded with gravitons instead of leptons. For unpolarised scattering the hadronic tensor is written as:
\begin{equation}
 W_{\mu\nu\alpha\beta}=\sum_{i=1}^3 F^G_iA^i_{\mu\nu\alpha\beta}.
\end{equation}
The tensors $A^i_{\mu\nu\alpha\beta}$ are constructed using the following quantities which are orthogonal to the momentum of the graviton, $q$, so that the conservation condition is automatically satisfied:
\begin{eqnarray}
\overline{P}_{\mu}&=&P_{\mu}-(P\cdot q)q_{\mu}/q^2,\\
\overline{g}_{\mu\nu}&=&g_{\mu\nu}-q_{\mu}q_{\nu}/q^2,\\
 \pi_{\mu\nu}&=&\overline{P}_{\mu}\overline{P}_{\nu}-\frac{1}{(d-1)}\overline{g}_{\mu\nu}\overline{P}^2,
\label{tensor}
\end{eqnarray}with $P_{\mu}$ the proton momentum.
The relevant tensors are
\begin{eqnarray}
 A^{(1)}_{\mu\nu\alpha\beta}&=&\pi_{\mu\nu}\pi_{\alpha\beta},\\\nonumber
A^{(2)}_{\mu\nu\alpha\beta}&=&\overline{P}_{\mu}\overline{P}_{\alpha}\overline{g}_{\nu\beta}+\overline{P}_{\nu}\overline{P}_{\alpha}\overline{g}_{\mu\beta}+\overline{P}_{\mu}\overline{P}_{\beta}\overline{g}_{\nu\alpha}+\overline{P}_{\nu}\overline{P}_{\beta}\overline{g}_{\mu\alpha}\\
&&-\frac{4}{d-1}(\overline{P}_{\mu}\overline{P}_{\nu}\overline{g}_{\alpha\beta}+\overline{P}_{\alpha}\overline{P}_{\beta}\overline{g}_{\mu\nu})+\frac{4}{(d-1)^2}\overline{P}^2\overline{g}_{\mu\nu}\overline{g}_{\alpha\beta},\\
A^{(3)}_{\mu\nu\alpha\beta}&=&\overline{g}_{\mu\alpha}\overline{g}_{\nu\beta}+\overline{g}_{\mu\beta}\overline{g}_{\nu\alpha}-\frac{2}{d-1}\overline{g}_{\mu\nu}\overline{g}_{\alpha\beta}.
\end{eqnarray}Each of the tensors above is traceless and orthogonal to the momentum transfer $q$, two conditions that need to be satisfied by the gravitational current. 
Linear combinations of the structure functions can be obtained by acting on the hadronic tensor with appropriate projections. We use as a basis for our projections the following set of tensors:
\begin{eqnarray}
P_{A}&=&g_{\nu\alpha}g_{\nu\beta}+g_{\mu\beta}g_{\nu\alpha},\\
P_{B}&=&P_{\alpha}P_{\beta}P_{\nu}P_{\mu},\\
P_{C}&=&g_{\mu\alpha}P_{\nu}P_{\beta}+g_{\nu\alpha}P_{\mu}P_{\beta}+g_{\mu\beta}P_{\nu}P_{\alpha}+g_{\nu\beta}P_{\mu}P_{\alpha}.
\end{eqnarray}
To isolate $F_1, F_2$ and $F_3$ we use the appropriate linear combinations of $P_A, P_B$ and $P_C$ which we write down here for simplicity in 4 dimensions, keeping in mind that d-dimensional expressions are needed when the NLO corrections are considered within dimensional regularisation:
\begin{eqnarray}
 P_1&=&\frac{2x^4}{Q^8}(Q^2P_C-20x^2P_B+560x^4Q^4P_A),\\
P_2&=&-\frac{16x^2}{Q^6}(Q^2P_C-2x^2P_B+20x^4Q^4P_A),\\
P_3&=&\frac{1}{2Q^4}(Q^2P_C-x^2P_B+2x^4Q^4P_A).
\end{eqnarray}
The expression for the double differential cross section in Eq.~\ref{LOgrav} is then written in terms of the structure functions defined above as
\begin{equation}
 \frac{d^2\sigma}{dQ^2dx}=\frac{\pi y^2Q^4}{8M^8_s}\bigg[\frac{1}{y^4x^4}(4-8y+5y^2-y^3)F_1-\frac{2}{y^2x^2}(2-2y+y^2)F_2-8F_3\bigg],
\end{equation}
where the expressions accompanying the structure functions are obtained by contracting the tensors $A^i_{\mu\nu\alpha\beta}$ with the leptonic tensor, which only depends on the lepton momentum and the momentum transfer. Note that we have rescaled the projection factors so that the structure functions are dimensionless. At leading order QCD, taking into account only the quark contributions, we have
\begin{eqnarray}\label{fq1}
F^G_1=2x^4[q(x)+\overline{q}(x)],\\ \label{fq2}
F^G_2=\frac{-x^2}{8}[q(x)+\overline{q}(x)],
\end{eqnarray}
 while $F^G_3$ has no quark contribution at LO.

The corresponding expression for gluon scattering is given by
\begin{equation}
 \frac{d^2\sigma}{dQ^2dx}=\frac{\pi}{2M^8_s}\frac{Q^4}{y^2}(2-4y+3y^2-y^3)g(x),
\end{equation}
which rewritten as above in terms of the structure functions, leads to the following gluon contributions to the structure functions:
\begin{eqnarray}\label{fg1}
F^G_1&=&2x^4g(x),\\ \label{fg2}
F^G_2&=&\frac{-x^2}{2}g(x),\\ 
F^G_3&=&\frac{1}{8}g(x).
\end{eqnarray}Adding all terms together, we note that at LO $F_1$ and $F_2$ receive contributions from both quark and gluon scattering, while $F_3$ only involves a gluonic contribution. 
\subsection{Sum rules}
In this subsection we comment on the existence of sum rules similar to those encountered for photon DIS. 
For photon DIS the Callan-Gross relation $F_2=2xF_1$ is satisfied at LO. This sum rule is explained by considering the longitudinal structure function $F_L=F_2-2xF_1$, which is zero at LO as a spin-1/2 quark cannot absorb a longitudinally polarised vector boson due to angular momentum conservation. It can be derived by using the polarisation vectors of the virtual boson to project out from the hadronic tensor the linear combination of $F_1$ and $F_2$ corresponding to the scattering of a longitudinal and transverse photon. At NLO the Callan-Gross relation is violated and $F_L=\mathcal{O}(\alpha_s)$. 

For the graviton, the derivation of a such sum rule at LO is not so straightforward, as now both spin-1/2 quarks and spin-1 gluons couple to the graviton. If we now use the polarisation tensors of the graviton, formed using polarisation vectors of massive gauge bosons \cite{Han:1998sg}, we find that the corresponding longitudinal structure function is a linear combination of $F^G_1$, $F^G_2$ and $F^G_3$ for which the quark contribution exactly vanishes, that is $F^G_L\propto F^G_1+16F^G_2$ expected from the same arguments that require $F^{\gamma}_L$ to be zero. Similarly by studying angular momentum conservation for the process in the Breit frame we see that a quark cannot absorb a graviton with spin component $\pm$2. Therefore the structure function corresponding to scattering of a helicity $\pm2$ graviton involves the same combination of $F^G_1$ and $F^G_2$, so that it receives no quark contribution and also a term proportional to $F^G_3$. Gluons can absorb any of the five possible polarisations of the graviton so there is no sum rule constraining the structure functions based on their gluonic parts. 
\section{Graviton exchange NLO results}
\subsection{Quark constituent contribution}
\begin{figure}[ht]
\begin{minipage}[b]{0.5\linewidth}
\centering
\includegraphics[scale=0.5]{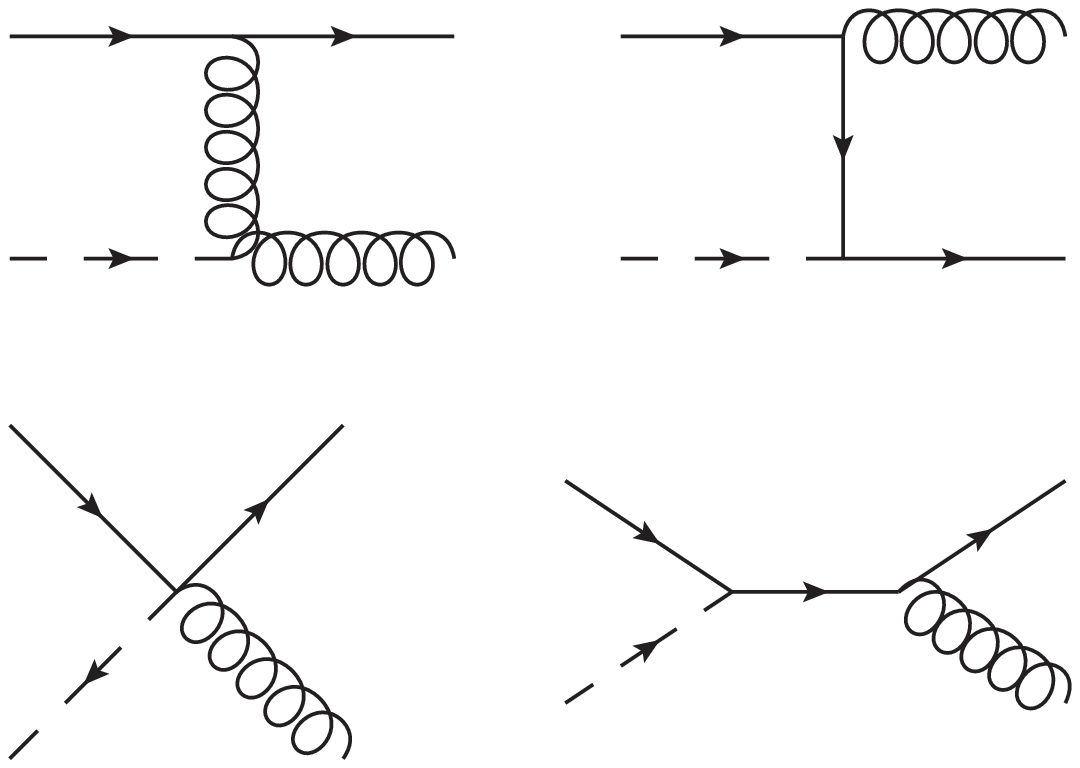}
\caption{Diagrams used to extract the real divergences.}
\label{sudgr}
\end{minipage}
\hspace{0.5cm}
\begin{minipage}[b]{0.5\linewidth}
\centering
\includegraphics[scale=0.5]{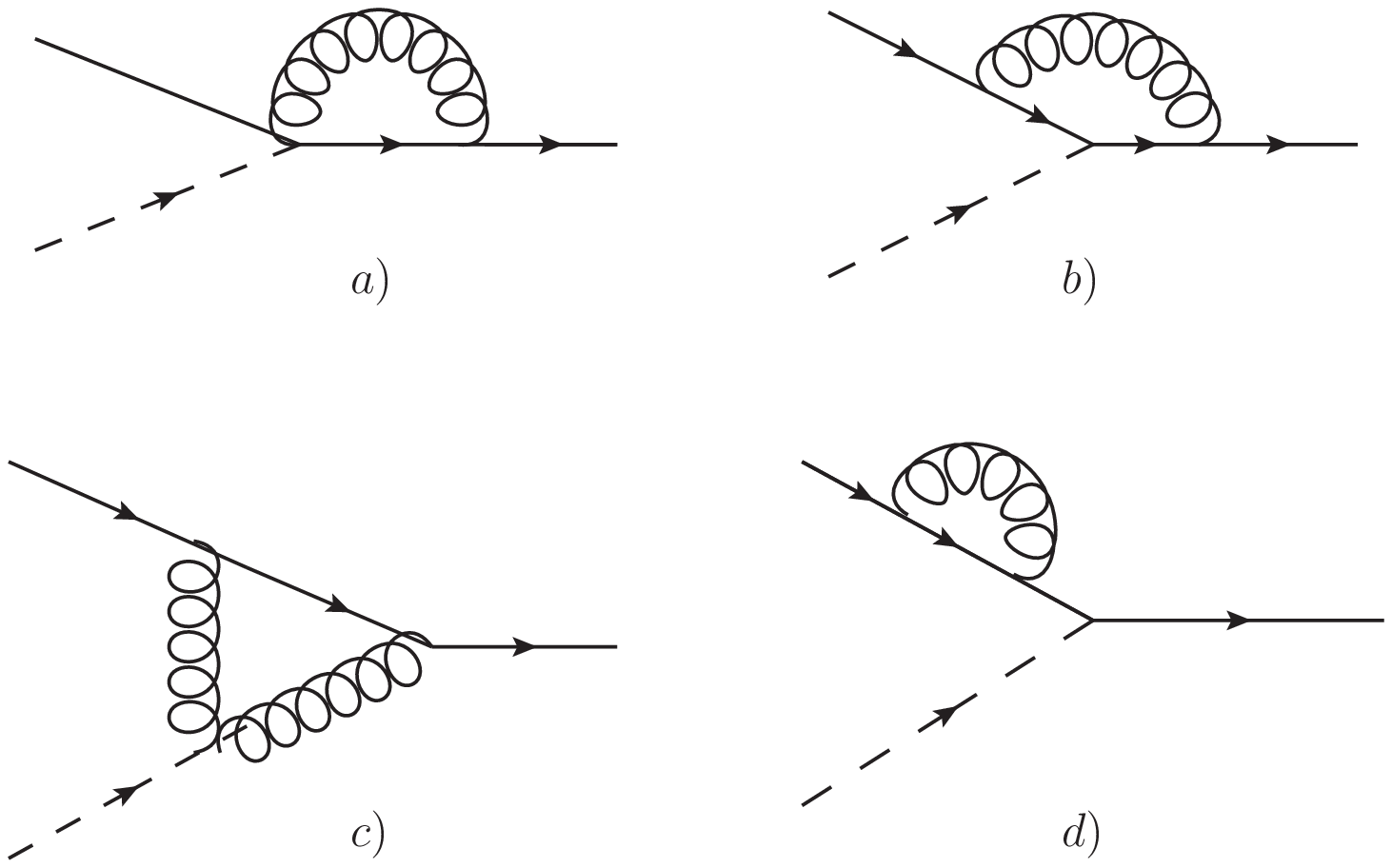}
\caption{Loop correction diagrams.}
\label{virtual}
\end{minipage}
\end{figure}

Reformulating the cross section in terms of structure functions allows us to extract analytic NLO results by simply projecting the structure functions using the NLO hadronic tensor.
This involves calculating the real and virtual correction diagrams. We start by considering initial state quark interactions. The diagrams used to extract the real corrections are shown in Fig.~\ref{sudgr} and the virtual loop diagrams in Fig.~\ref{virtual}. The calculation of the real emission diagrams is first performed at parton level, parameterising the momenta in terms of Sudakov variables.
For the $\mathcal{O}(\alpha_s)$ real corrections to the partonic structure functions we obtain (with $d=4-2\epsilon$)
\begin{eqnarray}\nonumber
\hat{F}^G_{1r}&=&\frac{C_F\alpha_s}{2\pi}\frac{(4\pi\mu^2/ Q^2)^{\epsilon}}{\Gamma(1-\epsilon)}2x^4\bigg[\frac{2}{\epsilon^2}\delta(1-x)-\frac{1}{\epsilon}\bigg(\frac{1+x^2}{1-x}\bigg)_+-\frac{1}{\epsilon}\frac{1+(1-x)^2}{x}+\frac{3}{\epsilon}\delta(1-x)\\\nonumber
&+&\big(\frac{7}{2}-\frac{\pi^2}{3}\big)\delta(1-x)+\bigg(\frac{\text{ln}(1-x)}{1-x}\bigg)_+\frac{2-3x+3x^2}{x}-\frac{1}{2x}\frac{1}{(1-x)_+}(3-20x+20x^2)\\
&-&\frac{1}{(1-x)_+}\frac{2-3x+3x^2}{x}\text{ln}x\bigg], 
\end{eqnarray}
\begin{eqnarray}\nonumber
\hat{F}^G_{2r}&=&-\frac{C_F\alpha_s}{2\pi}\frac{(4\pi\mu^2/Q^2)^{\epsilon}}{\Gamma(1-\epsilon)}\frac{x^2}{8}\bigg[\frac{2}{\epsilon^2}\delta(1-x)-\frac{1}{\epsilon}\bigg(\frac{1+x^2}{1-x}\bigg)_+-\frac{4}{\epsilon}\frac{1+(1-x)^2}{x}+\frac{3}{\epsilon}\delta(1-x)\\\nonumber
&+&(\frac{7}{2}-\frac{\pi^2}{3})\delta(1-x)-\bigg(\frac{\text{ln}(1-x)}{1-x}\bigg)_+\frac{-8+15x-12x^2+3x^3}{x}\\
&+&\frac{(-28+79x-74x^2+20x^3)}{2x}\frac{1}{(1-x)_+}+\frac{1}{(1-x)_+}\frac{-8+15x-12x^2+3x^3}{x}\text{ln}x\bigg], 
\end{eqnarray}
\begin{equation}
\hat{F}^G_{3r}=\frac{C_F\alpha_s}{2\pi}\frac{(4\pi\mu^2/Q^2)^{\epsilon}}{\Gamma(1-\epsilon)}\frac{1}{16}\bigg[-\frac{2}{\epsilon}\frac{1+(1-x)^2}{x}-\frac{7-10x+3x^2}{x}+4\text{ln}\bigg(\frac{(1-x)}{x}\bigg)\frac{(2-2x+x^2)}{x}\bigg],
\end{equation}with $\mu$ the renormalisation scale.
We note that here we write down only the order $\alpha_s$ piece of each structure function which has to be added to the LO results presented in the previous section. The results above show the expected splitting functions $P_{q\rightarrow q}$ and $P_{q\rightarrow g}$~\cite{Altarelli:1977zs} multiplying the collinear singularities as either of the quark or gluon decay products of the 
initial-state quark splitting can participate in the hard process. We also note that the relative prefactor of the splitting functions $P_{q\rightarrow q}$ and $P_{q\rightarrow g}$  differs between $F_1$ and $F_2$. This is directly related to the relative factor of quark and gluon contributions to $F_1$ and $F_2$ at LO as shown in Eqs.~\ref{fq1}, \ref{fq2}, \ref{fg1}, \ref{fg2}. The absence of the splitting function $P_{q\rightarrow q}$ from $F_3$ is explained again by the observation that at LO there is no quark contribution to $F_3$.

From the virtual corrections we obtain
\begin{eqnarray}
 \hat{F}^G_{1v}&=&\frac{C_F\alpha_s}{2\pi}\frac{(4\pi\mu^2/Q^2)^{\epsilon}}{\Gamma(1-\epsilon)}2x^4\left[-\frac{2}{\epsilon^2}-\frac{3}{\epsilon}-10\right]\delta(1-x),
\end{eqnarray}with diagrams a) and d) of Fig.~\ref{virtual} not contributing within dimensional regularisation. For $F_2$ the relative virtual corrections are the same as for $F_1$, while $F_3$ receives no contribution from the loop diagrams. Adding the virtual and real contributions demonstrates the expected cancellation of the IR divergences. The divergent terms proportional to the splitting functions which correspond to collinear divergences are absorbed into the PDFs. Within the $\overline{\text{MS}}$~\cite{Bardeen:1978yd} scheme the divergent term,
\begin{equation}
-\frac{1}{\epsilon}\frac{\alpha_s}{2\pi}\frac{(4\pi\mu_r^2/\mu_f^2)^{\epsilon}}{\Gamma(1-\epsilon)}P_{q\rightarrow q} \hat{F}_1^{LO}=(-\frac{1}{\epsilon}+\gamma-\text{log}(4\pi))\frac{\alpha_s}{2\pi} P_{q\rightarrow q}\hat{F}_1^{LO}
\end{equation} where $\hat{F}_1^{LO}$ is the LO partonic structure function at $\mathcal{O}(\epsilon)$,
 is absorbed into the PDF redefinition, with the normalisation and factorisation scale here set equal to $Q$. The form of this $\overline{\text{MS}}$ counter term is universal for all processes considered in the remainder of the calculation with the splitting function and a possible factor of 2 depending on the process.

To obtain the hadronic structure functions from the partonic results, we take the coefficient functions $C(x)$, defined for example for $\hat{F}_1$: 
\begin{equation}
\hat{F}_1=x^4\frac{\alpha_s}{2\pi}C^{\overline{\text{MS}}}_q(x),
\end{equation}
 change the argument to $z$, multiply by $q\big(\frac{x}{z}\big)/z$ and integrate over $z$ from $x$ to 1. 
For example the result we obtain for the $\alpha_s$ part of the NLO hadronic structure function $F_1$ involving the quark parton distribution functions is
\begin{eqnarray}\nonumber
F^G_{1}&=&\frac{C_F\alpha_s}{\pi}x^4\bigg[-(\frac{13}{2}+\frac{\pi^2}{3})q(x,Q^2)+\int_x^1\frac{dz}{z}q(\frac{x}{z},Q^2)\big[ \bigg(\frac{\text{ln}(1-z)}{1-z}\bigg)_+\frac{2-3z+3z^2}{z}\\
&-&\frac{(3-20z+20z^2)}{2z}\frac{1}{(1-z)_+}-\frac{\text{log}z}{(1-z)_+}\frac{2-3z+3z^2}{z}\big]\bigg]. 
\end{eqnarray}with $q(x,Q^2)$ the appropriate NLO parton distribution functions. Here there are no additional finite terms originating from considering the  LO structure functions at $\mathcal{O}(\epsilon)$, as with our choice of projections in d-dimensions only $F^g_2$ and $F^g_3$ acquire an additional factor of $(1+\epsilon)$, e.g. $F^g_2=-\frac{(1+\epsilon)}{2}x^2g(x)$. This has to be consistently taken into account in all the contributions to the structure functions that follow in order to extract the correct counter term. 
\subsection{Gluon constituent contribution}
\begin{figure}
\begin{minipage}[b]{0.5\linewidth}
\centering
\includegraphics[scale=0.6]{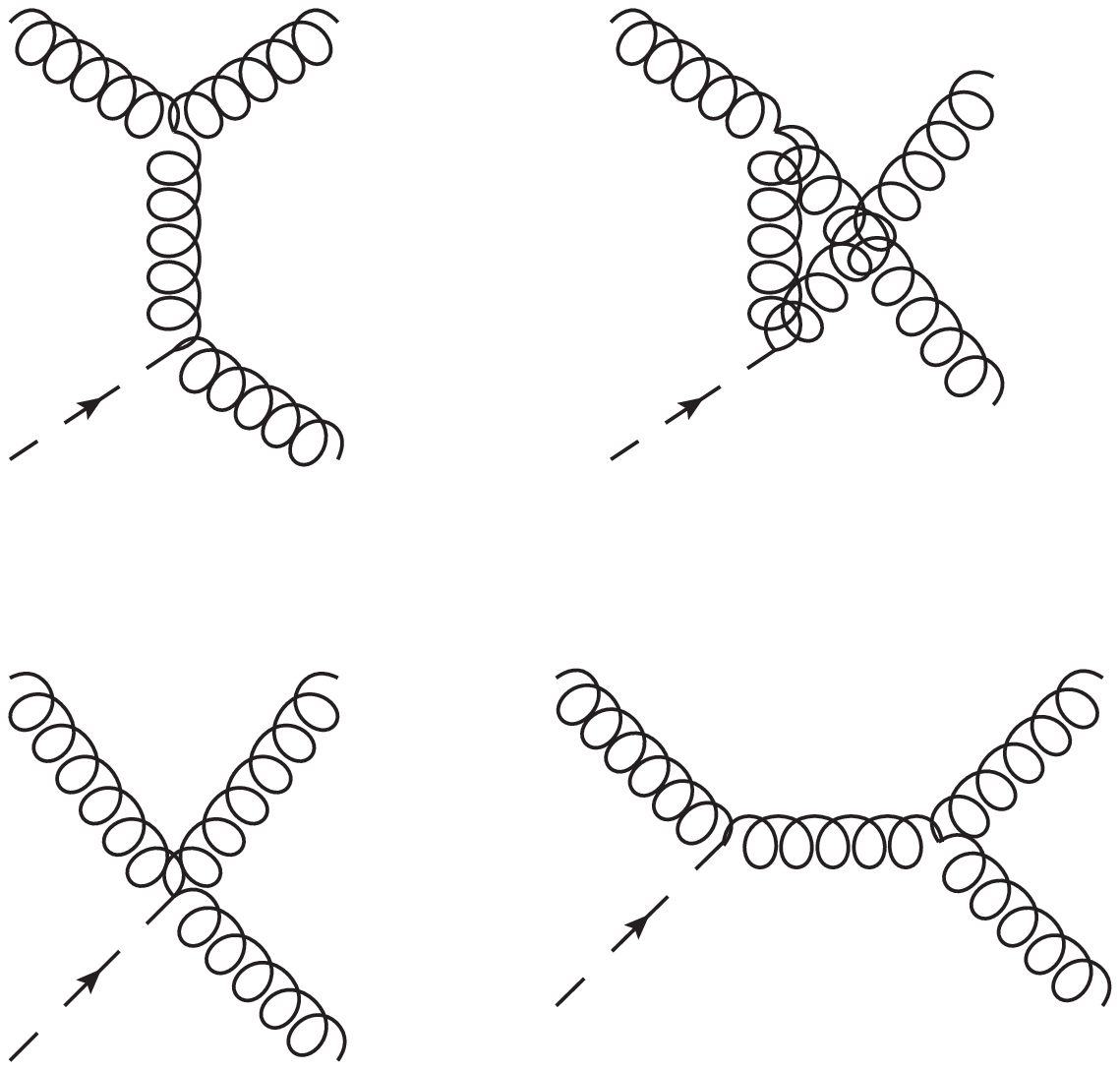}
\caption{Gluon constituent diagrams $gG^*\rightarrow gg$.}
\label{glugg}
\end{minipage}
\hspace{0.5cm}
\begin{minipage}[b]{0.5\linewidth}
\centering
\includegraphics[scale=0.6]{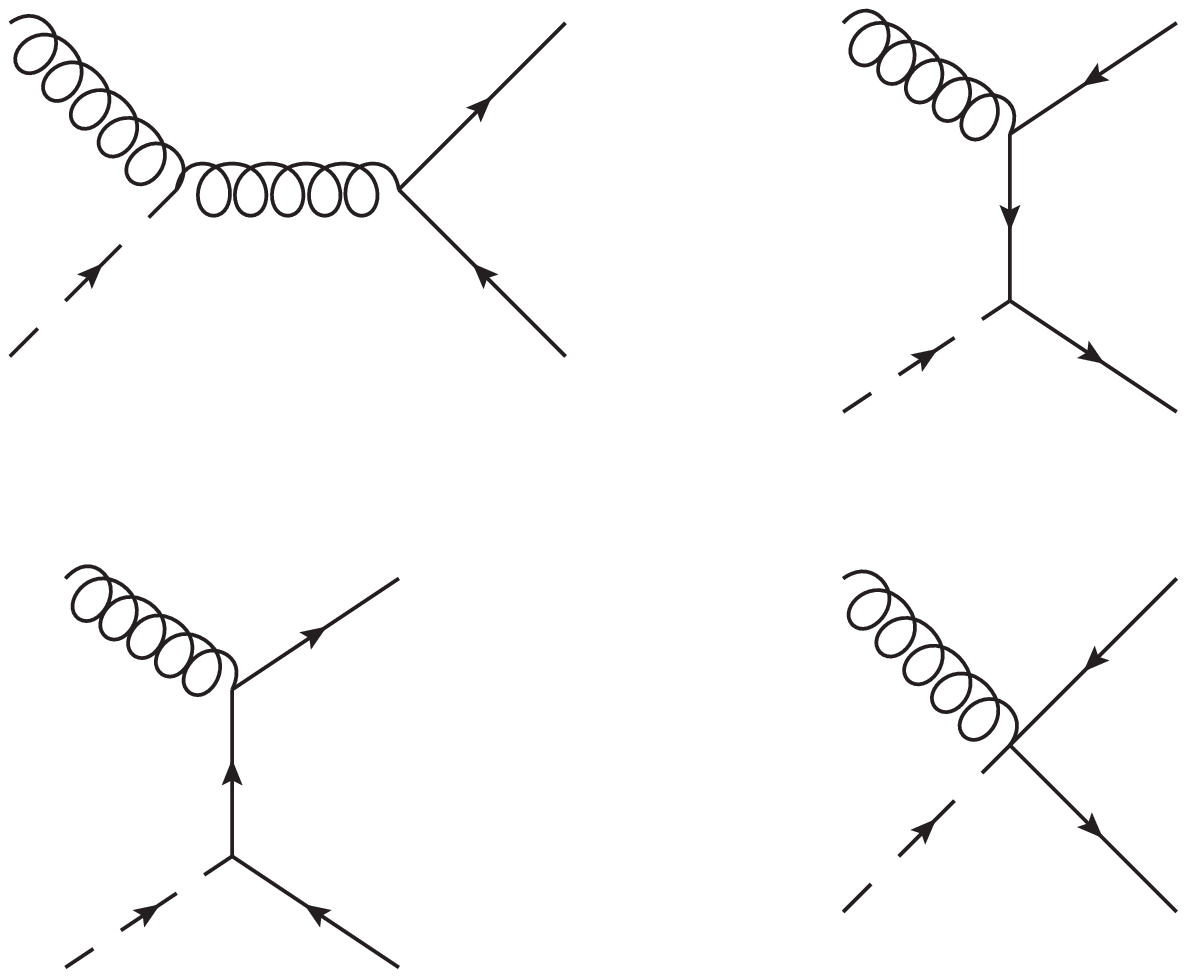}
\caption{Gluon constituent diagrams $gG^*\rightarrow q\overline{q}$.}
\label{gluonqq}
\end{minipage}
\end{figure}

\begin{figure}[ht]
\centering
\includegraphics[scale=0.4,angle=-90]{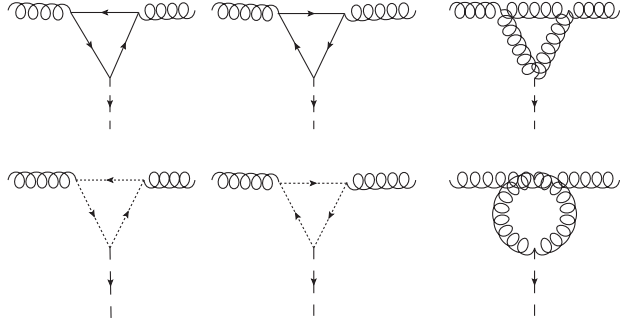}
\caption{Initial state gluon virtual correction diagrams.}
\label{glu}
\end{figure}

The same procedure is followed to calculate the gluon contributions to the structure functions at NLO. The relevant Feynman diagrams for the real corrections are shown in Fig.~\ref{glugg} and \ref{gluonqq}, while the virtual diagrams are shown in Fig.~\ref{glu}, with only the diagrams non-vanishing within dimensional regularisation shown. For the real corrections, there exist two sets of diagrams with different final states. We note that for the diagrams in Fig.~\ref{glugg} we have two identical particles in the final states, which implies that we need to divide by two to  avoid double counting. For all matrix element calculations, when summing over the polarisation of external gluons we use 
\begin{eqnarray}
\sum_T \epsilon_{T}^{\mu*}(k)\epsilon_{T}^{\nu}(k)=-g^{\mu\nu}+\frac{k^{\mu}n^{\nu}+k^{\nu}n^{\mu}}{nk}-n^2\frac{k^{\mu}k^{\nu}}{(nk)^2},
\end{eqnarray}
in order to include only the physical gluon polarisations. In the results that follow we take into account the additional factor of $\frac{1}{1-\epsilon}$ coming from averaging over the initial gluon polarisations in d-dimensions. This factor will result in a finite term proportional to the splitting function remaining in the coefficient function.

The results for the partonic structure functions from the diagrams of Fig.~\ref{glugg} are
\begin{eqnarray}\nonumber
 \hat{F}^G_{1r}&=&\frac{3\alpha_s}{\pi}\frac{(4\pi\mu^2/Q^2)^{\epsilon}}{(1-\epsilon)\Gamma(1-\epsilon)}x^4\bigg[\frac{2\delta(1-x)}{\epsilon^2}-2\frac{1}{(1-x)_+}\frac{(x^2-x+1)^2}{x\epsilon}-\frac{1}{6\epsilon}\delta(1-x)\\\nonumber
&-&\frac{1}{(1-x)_+}\frac{(38x^4-76x^3+90x^2-52x-1)}{6x}+2\bigg(\frac{\text{ln}(1-x)}{1-x}\bigg)_+\frac{(x^2-x+1)^2}{x}\\
&+&\frac{8}{9}\delta(1-x)-\frac{\pi^2}{3}\delta(1-x)-2\frac{1}{(1-x)_+}\frac{(x^2-x+1)^2}{x}\text{log}x\bigg]\\\nonumber
&=&\frac{3\alpha_s}{\pi}\frac{(4\pi\mu^2/Q^2)^{\epsilon}}{(1-\epsilon)\Gamma(1-\epsilon)}x^4\bigg[\frac{2\delta(1-x)}{\epsilon^2}-\frac{\text{P}_{g\rightarrow g}}{3\epsilon}-\frac{1}{18\epsilon}(-30+2n_f)\delta(1-x)\\\nonumber
&-&\frac{1}{(1-x)_+}\frac{(38x^4-76x^3+90x^2-52x-1)}{6x}+2\bigg(\frac{\text{ln}(1-x)}{1-x}\bigg)_+\frac{(x^2-x+1)^2}{x}\\
&+&\frac{8}{9}\delta(1-x)-\frac{\pi^2}{3}\delta(1-x)-2\frac{1}{(1-x)_+}\frac{(x^2-x+1)^2}{x}\text{log}x\bigg].
\end{eqnarray}
In the results for the real corrections we can identify the Altarelli-Parisi~\cite{Altarelli:1977zs} splitting function:
\begin{equation}
P_{g\rightarrow g}=6\left[\frac{z}{(1-z)_+}+\frac{1-z}{z}+z(1-z)\right]+\frac{1}{6}(33-2n_f)\delta(1-z)
\end{equation}by rearranging the terms, as 
\begin{equation}
\bigg(\frac{1}{1-z}\bigg)_+\frac{(1-z+z^2)^2}{z}=\left[\frac{z}{(1-z)_+}+\frac{1-z}{z}+z(1-z)\right].
\end{equation}
Similarly for $F_2$ and $F_3$,
\begin{eqnarray}\nonumber
 \hat{F}^G_{2r}&=&-\frac{3\alpha_s}{2\pi}\frac{(4\pi\mu^2/Q^2)^{\epsilon}}{(1-\epsilon)\Gamma(1-\epsilon)}\frac{x^2}{2}\bigg[\frac{2\delta(1-x)}{\epsilon^2}-2\frac{1}{(1-x)_+}\frac{(x^2-x+1)^2}{x\epsilon}+\frac{11}{6\epsilon}\delta(1-x)\\\nonumber
&-&\frac{1}{(1-x)_+}\frac{(38x^4-76x^3+90x^2-52x+11)}{6x}+2\bigg(\frac{\text{ln}(1-x)}{1-x}\bigg)_+\frac{(x^2-x+1)^2}{x}\\
&+&\frac{85}{18}\delta(1-x)-\frac{\pi^2}{3}\delta(1-x)-2\frac{1}{(1-x)_+}\frac{(x^2-x+1)^2}{x}\text{log}x\bigg],
\end{eqnarray}
\begin{eqnarray}\nonumber
 \hat{F}^G_{3r}&=&\frac{3\alpha_s}{2\pi}\frac{(4\pi\mu^2/Q^2)^{\epsilon}}{(1-\epsilon)\Gamma(1-\epsilon)}\frac{1}{8}\bigg[\frac{2\delta(1-x)}{\epsilon^2}-2\frac{1}{(1-x)_+}\frac{(x^2-x+1)^2}{x\epsilon}+\frac{11}{6\epsilon}\delta(1-x)\\\nonumber
&+&\frac{1}{(1-x)_+}\frac{(-22x^4+44x^3-66x^2+44x-11)}{6x}+2\bigg(\frac{\text{ln}(1-x)}{1-x}\bigg)_+\frac{(x^2-x+1)^2}{x}\\
&+&\frac{85}{18}\delta(1-x)-\frac{\pi^2}{3}\delta(1-x)-2\frac{1}{(1-x)_+}\frac{(x^2-x+1)^2}{x}\text{log}x\bigg].
\end{eqnarray}
The corresponding gluon and ghost loop diagram results for Fig.~\ref{glu} are
\begin{eqnarray}
 \hat{F}^G_{1v}&=&\frac{3\alpha_s}{\pi}\frac{(4\pi\mu^2/Q^2)^{\epsilon}}{(1-\epsilon)\Gamma(1-\epsilon)}x^4\left[-\frac{2}{\epsilon^2}-\frac{5}{3\epsilon}-\frac{119}{18}\right]\delta(1-x),
\end{eqnarray}
\begin{eqnarray}
 \hat{F}^G_{2v}&=&-\frac{3\alpha_s}{2\pi}\frac{(4\pi\mu^2/Q^2)^{\epsilon}}{(1-\epsilon)\Gamma(1-\epsilon)}\frac{x^2}{2}\left[-\frac{2}{\epsilon^2}-\frac{11}{3\epsilon}-\frac{221}{18}\right]\delta(1-x),
\end{eqnarray}
\begin{eqnarray}
 \hat{F}^G_{3v}&=&\frac{3\alpha_s}{2\pi}\frac{(4\pi\mu^2/Q^2)^{\epsilon}}{(1-\epsilon)\Gamma(1-\epsilon)}\frac{1}{8}\left[-\frac{2}{\epsilon^2}-\frac{11}{3\epsilon}-\frac{221}{18}\right]\delta(1-x).
\end{eqnarray}

Rewriting all the results for the real corrections above, using the expression for the splitting function and adding the loop results, we see that we are only left with the collinear divergences and the finite terms. Calculating the LO structure function at $\mathcal{O}(\epsilon)$ to find the appropriate counterterm for the collinear divergence, we obtain the finite coefficient function.
Similarly from the diagrams of Fig.~\ref{gluonqq} we obtain
\begin{eqnarray}\nonumber
  \hat{F}^G_{1r}&=&\frac{n_f\alpha_s}{4\pi}\frac{(4\pi\mu^2/Q^2)^{\epsilon}}{(1-\epsilon)\Gamma(1-\epsilon)}x^4\bigg[-\frac{4}{3\epsilon}\delta(1-x)-\frac{4}{\epsilon}(2x^2-2x+1)-\frac{8}{9}\delta(1-x)\\\nonumber
&-&4\bigg(\frac{\text{ln}(1-x)}{1-x}\bigg)_+(-1+3x-4x^2+2x^3)+4(-1+2x-2x^2)\text{log}x\\
&+&\frac{4}{3x}\frac{1}{(1-x)_+}(1-8x+33x^2-50x^3+25x^4)\bigg],\\
\nonumber
\hat{F}^G_{2r}&=&-\frac{n_f\alpha_s}{4\pi}\frac{(4\pi\mu^2/Q^2)^{\epsilon}}{(1-\epsilon)\Gamma(1-\epsilon)}\frac{x^2}{4}\bigg[-\frac{4}{3\epsilon}\delta(1-x)-\frac{1}{\epsilon}(2x^2-2x+1)-\frac{20}{9}\delta(1-x)\\\nonumber
&-&\bigg(\frac{\text{ln}(1-x)}{1-x}\bigg)_+(-1+3x-4x^2+2x^3)+(-1+2x-2x^2)\text{log}x\\
&+&\frac{1}{3x}\frac{1}{(1-x)_+}(4-20x+42x^2-44x^3+22x^4)\bigg],\\\nonumber
\hat{F}^G_{3r}&=&\frac{n_f\alpha_s}{4\pi}\frac{(4\pi\mu^2/Q^2)^{\epsilon}}{(1-\epsilon)\Gamma(1-\epsilon)}\frac{1}{16}\bigg[-\frac{4\delta(1-x)}{3\epsilon}-\frac{20\delta(1-x)}{9}\\
&+&\frac{4(1-4x+6x^2-4x^3+2x^4)}{3x(1-x)_+}\bigg],
\end{eqnarray} where $n_f$ is the number of active flavours. In these expressions, we identify as expected the collinear divergence term proportional to the Altarelli-Parisi splitting function $P_{g\rightarrow q}$. We note that $F_3$ receives no collinear divergence contribution from the set of diagrams in Fig.~\ref{gluonqq}. This is expected, as at LO there is no quark contribution to $F_3$. A similar result is obtained for the gluonic contribution to $F_L$ for the electromagnetic current.
Similarly for the fermion loop diagrams of Fig.~\ref{glu} we have
\begin{eqnarray}
\hat{F}^G_{1v}&=&\frac{n_f\alpha_s}{2\pi}\frac{(4\pi\mu^2/Q^2)^{\epsilon}}{(1-\epsilon)\Gamma(1-\epsilon)}x^4\left[\frac{4}{3\epsilon}+\frac{23}{9}\right]\delta(1-x),\\
\hat{F}^G_{2v}&=&-\frac{n_f\alpha_s}{2\pi}\frac{(4\pi\mu^2/Q^2)^{\epsilon}}{(1-\epsilon)\Gamma(1-\epsilon)}\frac{x^2}{4}\left[\frac{4}{3\epsilon}+\frac{35}{9}\right]\delta(1-x),\\
\hat{F}^G_{3v}&=&\frac{n_f\alpha_s}{2\pi}\frac{(4\pi\mu^2/Q^2)^{\epsilon}}{(1-\epsilon)\Gamma(1-\epsilon)}\frac{1}{16}\left[\frac{4}{3\epsilon}+\frac{35}{9}\right]\delta(1-x).
\end{eqnarray}Combining the results proportional to the number of flavours from the loop diagrams and the real emission diagrams we verify that the single pole divergences vanish.
The results for the real and virtual correction diagrams have been obtained using FORM~\cite{Vermaseren:2000nd} and Mathematica, manually performing the Passarino-Veltman~\cite{Passarino:1978jh} reduction for the loop diagrams. As a check, a subset of the results were tested using the Mathematica package FeynCalc~\cite{Mertig:1990an} which automatically performs the Passarino-Veltman reduction.
\section{Interference with SM gauge bosons}
\begin{figure}[h]
\centering
\includegraphics[scale=0.4]{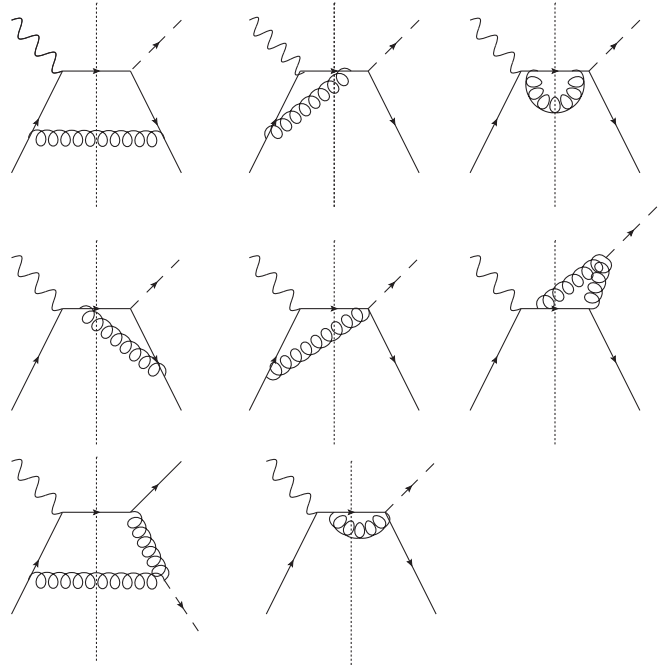}
\caption{Graviton-gauge boson interference diagrams.}
\label{intdiag}
\end{figure}

\begin{figure}[h]
\centering
\includegraphics[scale=0.4]{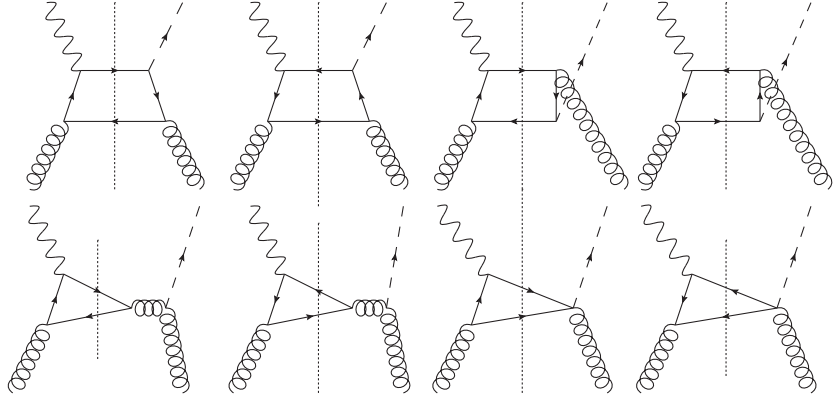}
\caption{Interference diagrams with initial state gluon.}
\label{interg}
\end{figure}

As the graviton is a colour singlet, interference effects with other colour singlet gauge bosons come into play when we consider the DIS cross section. For the initial state quark interactions we need to consider the interference with the photon and Z boson. At LO the $G\gamma$ interference for positron DIS gives the following contribution to the cross section:
\begin{equation}
 \frac{d\hat{\sigma}^{\gamma G}}{dQ^2}=\frac{-\lambda\pi\alpha e_q}{2M_S^4}\frac{8-12y+6y^2-y^3}{y},
\end{equation}while the $GZ$ interference is
\begin{equation}
 \frac{d\hat{\sigma}^{Z G}}{dQ^2}=\frac{\lambda\pi\alpha}{2M_S^4\rm{sin}^2 2\theta_w}\frac{Q^2}{(Q^2+M_Z^2)}\frac{[c^q_vc_v^e(8-12y+6y^2-y^3)-c^q_ac_a^e(6y-6y^2+y^3)]}{y},
\end{equation}where $\lambda=\pm 1$ and the sign determines if the interference is constructive or destructive.
By analogy with the pure photon and pure graviton contributions, the hadronic tensor will be of the form
\begin{equation}
 W_{\mu\rho\sigma}\propto \int d^4z\; e^{iqz} \langle P,S|T_{\rho\sigma}(z)j_{\mu}(0)|P,S\rangle, 
\end{equation}where $j_{\mu}$ is the electromagnetic current.
Using the quantities defined in Eq.~\ref{tensor} that are orthogonal to $q$, we form a set of three tensors which satisfy the conditions of trace and orthogonality,
\begin{eqnarray}
W_1^{\mu\rho\sigma}&=&\pi^{\rho\sigma}\overline{P}^{\mu},\\
W_2^{\mu\rho\sigma}&=&\overline{g}^{\mu\sigma}\overline{P}^{\rho}+\overline{g}^{\mu\rho}\overline{P}^{\sigma}-\frac{2}{3}\overline{g}^{\rho\sigma}\overline{P}^{\mu},\\
W_3^{\mu\rho\sigma}&=&\epsilon^{\rho\mu\alpha\beta}\overline{P}^{\sigma}\overline{P}_{\alpha}q_{\beta}+\epsilon^{\sigma\mu\alpha\beta}\overline{P}^{\rho}\overline{P}_{\alpha}q_{\beta}.
\end{eqnarray}
For the graviton-photon interference only the first two are relevant.
These are slightly modified for $d\neq 4$ to ensure the trace condition is still satisfied, with special care needed for the treatment of the 4-dimensional $\epsilon$ tensor. When contracted with the leptonic tensor, taking into account only the Dirac matrices structure, we obtain the following functions of $x,y$ and $Q^2$:
\begin{eqnarray}
\frac{Q^6}{x^3y^3}(8-12y+4y^2),\\
\frac{8Q^4}{xy^3}(-2y^2+y^3),\\
\frac{4Q^6 c_a^e}{x^2y^3}(6y-6y^2+y^3).
\end{eqnarray}
To extract the structure functions we use linear combinations of the following projections:
\begin{eqnarray}
P_A=P_{\sigma}P_{\rho}P_{\mu} \text{     and   }&& P_B=g_{\mu\rho}P_{\sigma}+g_{\mu\sigma}P_{\rho},
\end{eqnarray}with $F_1$ and $F_2$ obtained from
\begin{eqnarray}
P_1=-\frac{8x^4}{Q^6}(Q^2 P_B-20x^2P_A)\ \text{     and   }&& P_2=\frac{x^2}{Q^4}(Q^2P_B-8x^2P_A).
\end{eqnarray}
We can therefore rewrite the differential cross section in the form
\begin{equation}
\frac{d\sigma^{\gamma G}}{dxdQ^2}=\frac{-\lambda\pi\alpha e_q y^2}{2M_S^4}[\frac{(8-12y+4y^2)}{y^3x^2}F_1+8\frac{(-2y^2+y^3)}{y^3}F_2],
\end{equation}
which at LO gives
\begin{eqnarray}
F^{\gamma G}_1=x^2 [q(x)-\overline{q}(x)] &&\text{and  } F^{\gamma G}_2=-\frac{1}{8}[q(x)-\overline{q}(x)],
\end{eqnarray}where once more the structure functions are defined to be dimensionless. The quark and antiquark contributions to both $F^{\gamma G}_1$ and $F^{\gamma G}_2$ have opposite signs due to the charge factor present in the expression.
To make the calculation slightly clearer, we can redefine the tensor $W_1$ and therefore $F_1^{\gamma G}$ by moving the $x^2$ factor of $F^{\gamma G}_1$ to the tensor $W_1$ which leaves the LO partonic $F^{\gamma G}_1$ with no explicit factor of $x$. At the partonic level and again using Sudakov parameterisation for the momenta we calculate the diagrams of Fig.~\ref{intdiag}, where the dotted lines denote the final state cut and find
\begin{eqnarray}\nonumber
\hat{F}^{\gamma G}_1&=& \frac{C_F\alpha_s}{2\pi}\frac{(4\pi\mu^2/Q^2)^{\epsilon}}{\Gamma(1-\epsilon)}\bigg[\frac{2}{\epsilon^2}\delta(1-x)-\frac{1}{(1-x)_+}\frac{(1+x^2)}{\epsilon}+\frac{3}{2\epsilon}\delta(1-x)\\\nonumber
&+&\frac{7}{2}\delta(1-x)-\frac{\pi^2}{3}\delta(1-x)+\bigg(\frac{\text{ln}(1-x)}{1-x}\bigg)_+(1+x^2)\\
&-&\frac{1}{2}\frac{1}{(1-x)_+}(-5+4x+4x^2)-\frac{1}{(1-x)_+}(1+x^2)\text{ln}x\bigg].
\end{eqnarray}Similarly for $\hat{F}_2$ we obtain
\begin{eqnarray}\nonumber
\hat{F}^{\gamma G}_2&=&-\frac{C_F\alpha_s}{2\pi}\frac{(4\pi\mu^2/Q^2)^{\epsilon}}{\Gamma(1-\epsilon)}\frac{1}{8}\bigg[\frac{2}{\epsilon^2}\delta(1-x)-\frac{1}{(1-x)_+}\frac{(1+x^2)}{\epsilon}+\frac{3}{2\epsilon}\delta(1-x)\\\nonumber
&+&\frac{7}{2}\delta(1-x)-\frac{\pi^2}{3}\delta(1-x)+\bigg(\frac{\text{ln}(1-x)}{1-x}\bigg)_+(1+x^2)\\
&-&\frac{1}{2}\frac{1}{(1-x)_+}(-5+10x-2x^2)-\frac{1}{(1-x)_+}(1+x^2)\text{ln} x\bigg].
\end{eqnarray}We note that apart from an overall factor, the two expressions differ only in the finite terms.
Calculating the interference terms between a graviton exchange diagram at $\mathcal{O}(\alpha_s)$ and a LO order photon exchange diagram and vice versa we obtain the following expression for the virtual corrections:
\begin{eqnarray}
\hat{F}^{\gamma G}_{1v}&=& \frac{C_F\alpha_s}{2\pi}\frac{(4\pi\mu^2/Q^2)^{\epsilon}}{\Gamma(1-\epsilon)}\bigg[-\frac{2}{\epsilon^2}-\frac{3}{\epsilon}-9\bigg]\delta(1-x) .
\end{eqnarray}The constant term originates from combining the constant term contribution from the virtual loop in photon exchange ($-8$) and graviton exchange ($-10$). The result for $\hat{F}_2^{\gamma G}$ is identical. Using
\begin{equation}
 (1+x^2)\frac{1}{(1-x)_+}=\bigg(\frac{1+x^2}{1-x}\bigg)_+-\frac{3}{2}\delta(1-x),
\end{equation}
we notice that as expected the double and single pole divergences exactly cancel when we add the virtual and real contributions, and that we obtain the expected form of the collinear divergence:
\begin{equation}
-\frac{C_F\alpha_s}{2\pi}\frac{(4\pi\mu_r^2/\mu_f^2)^{\epsilon}}{\Gamma(1-\epsilon)}\frac{1}{\epsilon}P_{q\rightarrow q}.
\end{equation}
The relevant coefficient function is
\begin{eqnarray}\nonumber
\hat{F}^{\gamma G}_1&=& \frac{C_F\alpha_s}{2\pi}\bigg[+\frac{7}{2}\delta(1-x)-9\delta(1-x)-\frac{\pi^2}{3}\delta(1-x)+\bigg(\frac{\text{ln}(1-x)}{1-x}\bigg)_+(1+x^2)\\
&-&\frac{1}{2}\frac{1}{(1-x)_+}(-5+4x+4x^2)-\frac{1}{(1-x)_+}(1+x^2)\text{ln}x\bigg].
\end{eqnarray}
Using
\begin{equation}
 \bigg(\frac{\text{ln}(1-x)}{1-x}\bigg)_+(1+x^2)=2\bigg(\frac{\text{ln}(1-x)}{1-x}\bigg)_+-(1+x)\text{ln}(1-x)
\end{equation}we obtain
\begin{eqnarray}\nonumber
\hat{F}^{\gamma G}_1&=& \frac{C_F\alpha_s}{2\pi}\bigg[2\bigg(\frac{\text{ln}(1-x)}{1-x}\bigg)_+-\frac{11}{2}\delta(1-x)-\frac{\pi^2}{3}\delta(1-x)-(1+x)\text{ln} (1-x)\\
&-&\frac{1}{2}\frac{1}{(1-x)_+}(-5+4x+4x^2)-\frac{1}{(1-x)_+}(1+x^2)\text{ln}x\bigg],
\end{eqnarray}
and correspondingly for $\hat{F}_2$:
\begin{eqnarray}\nonumber
\hat{F}^{\gamma G}_2&=&-\frac{1}{8} \frac{C_F\alpha_s}{2\pi}\bigg[2\bigg(\frac{\text{ln}(1-x)}{1-x}\bigg)_+-\frac{11}{2}\delta(1-x)-\frac{\pi^2}{3}\delta(1-x)-(1+x)\text{ln} (1-x)\\
&-&\frac{1}{2}\frac{1}{(1-x)_+}(-5+10x-2x^2)-\frac{1}{(1-x)_+}(1+x^2)\text{ln}x\bigg].
\end{eqnarray}
We note that there is a second set of interference diagrams with initial state gluons shown in Fig.~\ref{interg}. We calculate the matrix element squared for this contribution; however, the result is odd in $\text{cos}\theta^*$ where $\theta^*$ is the angle between the outgoing quark and the initial state gluon in the gluon-graviton(/photon) centre-of-mass frame. Therefore when we perform the two-body phase space integration we get no contribution to the structure functions $F_1^{\gamma G}$ and $F_2^{\gamma G}$. A similar effect arises in \cite{Mathews:2004xp} where the authors note that there is no interference between the graviton and SM contribution to Drell-Yan. This does not occur at the matrix element squared level but only when the phase space integration is performed. 

For the Z-graviton interference the odd parity structure function comes into play. The projection used to extract it is
\begin{equation}
 P_3=-\frac{4x^4}{Q^6}(\epsilon_{\sigma\mu\beta\kappa}P^{\beta}P^{\kappa}P_{\rho}+\epsilon_{\rho\mu\beta\kappa}P^{\beta}P^{\kappa}P_{\sigma}),
\end{equation}with the other two projections remaining the same. The differential cross section is now written as
\begin{eqnarray}
\frac{d\sigma^{Z G}}{dxdQ^2}& =& \frac{\pi\alpha\lambda}{2M_S^4\rm{sin}^2 2\theta_w}\frac{Q^2 y^2}{(Q^2+M_Z^2)}\left[ 
c_v^e\frac{(8-12y+4y^2)}{y^3x^2}F_1 \right. \nonumber \\
& & \left. +8c_v^e\frac{(-2y^2+y^3)}{y^3}F_2-4c_a^e\frac{(6y-6y^2+y^3)}{xy^3}F_3\right], 
\end{eqnarray}
where at LO
\begin{eqnarray}
F^{GZ}_1&=&x^2 c^q_v[q(x)-\overline{q}(x)],\\
F^{GZ}_2&=&-\frac{c^q_v}{8}[q(x)-\overline{q}(x)],\\
F^{GZ}_3&=&\frac{c^q_ax}{4}[q(x)+\overline{q}(x)].
\end{eqnarray}
As the first two structure functions are independent of $\gamma^5$ and given the universality of the $\mathcal{O}(\alpha_s)$ QCD corrections for spin-1 gauge bosons, we expect the NLO structure functions $F_1^{GZ}$ and $F_2^{GZ}$ to receive the same relative corrections as  $F_1^{\gamma G}$ and $F_2^{\gamma G}$ respectively. 
For the NLO contribution to $F_3^{GZ}$ we need to consistently treat $\gamma^5$ in $d$ dimensions. Following the prescription in \cite{Zijlstra:1992kj}, we replace the axial current by
\begin{equation}
 \overline{\psi}\gamma_{\mu}\gamma_5\psi=\frac{i}{6}\epsilon_{\mu\rho\sigma\tau}\overline{\psi}\gamma^{\rho}\gamma^{\sigma}\gamma^{\tau}\psi,
\end{equation} and then perform the trace with the gamma matrices in d-dimensions and keep the product of the two $\epsilon$ factors (one from the projection and one from the matrix element) outside the trace. The product of epsilon tensors gives a determinant of Kronecker deltas which are then treated as d-dimensional objects. Following consistently this prescription for the virtual and real corrections, we obtain for the real corrections to the partonic structure function $\hat{F}_3^{ZG}$
\begin{eqnarray}\nonumber
\hat{F}_3^{ZG}&=&\frac{C_F\alpha_sc^q_a x}{8\pi}\frac{(4\pi\mu^2/Q^2)^{\epsilon}}{\Gamma(1-\epsilon)}\bigg[\frac{2}{\epsilon^2}\delta(1-x)-\frac{1}{\epsilon}\bigg(\frac{1+x^2}{1-x}\bigg)_+ +\frac{1}{\epsilon}\delta(1-x)\\\nonumber
&+&(1+x^2)\bigg(\frac{\text{ln}(1-x)}{1-x}\bigg)_++2\delta(1-x)-\frac{\pi^2}{3}\delta(1-x)\\
&-&\frac{1}{2}\frac{1}{(1-x)_+}(2x^2-2x-1)-\text{log}x(1+x^2)\frac{1}{(1-x)_+}\bigg],
\end{eqnarray}while from combining the loop corrections to Z and graviton vertices we get for the virtual corrections
\begin{equation}
\hat{F}_3^{ZG}=\frac{c^q_aC_F\alpha_s}{8\pi}\frac{(4\pi\mu^2/Q^2)^{\epsilon}}{\Gamma(1-\epsilon)}\bigg[-\frac{2}{\epsilon^2}-\frac{1}{\epsilon}-4\bigg]\delta(1-x),
\end{equation}and therefore the divergences cancel as expected.
\section{Numerical results}
The analytic results for the cross sections are implemented into a program to numerically integrate over the momentum fraction and the DIS variable $x$.
The integration limits for $x$ are normally 0 to 1 with the lower limit modified to $Q^2/0.9s$ because experimentally at HERA the maximum value for $y$ is 0.9 \cite{Adloff:2003uh}. The proton energy is 920~GeV and the positron energy is 26.7~GeV and we calculate the differential cross section in the range of experimentally measured values of $Q^2$.

\begin{figure}[h]
\centering
\includegraphics[scale=0.6]{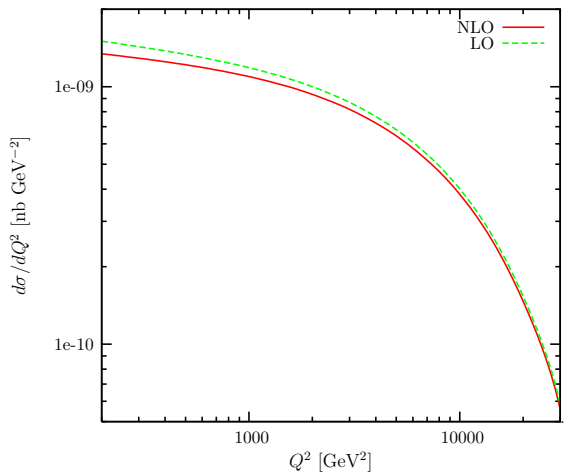}
\caption{Differential cross section for graviton exchange for $M_S=1$~TeV.}
\label{diffcro}
\end{figure}

\begin{figure}[h]
\begin{minipage}[b]{0.5\linewidth}
\centering
\includegraphics[trim=2.2cm 0 0 0,scale=0.6]{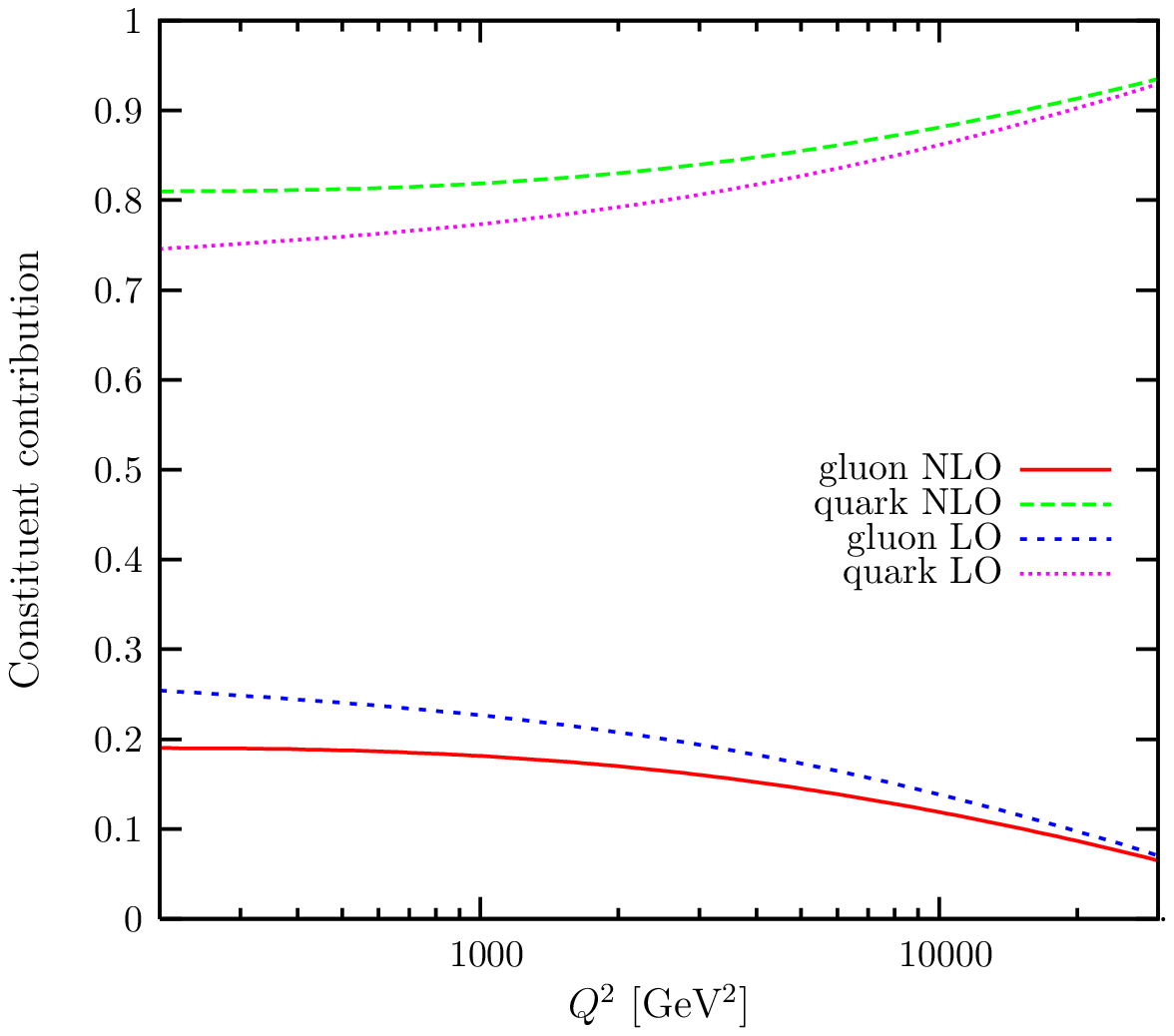}
\caption{Relative contributions to the cross section from initial state quarks and gluons.}
\label{decomp}
\end{minipage}
 \hspace{0.5cm}
\begin{minipage}[b]{0.5\linewidth}
\centering
\includegraphics[trim=1.6cm 0 0 0,scale=0.6]{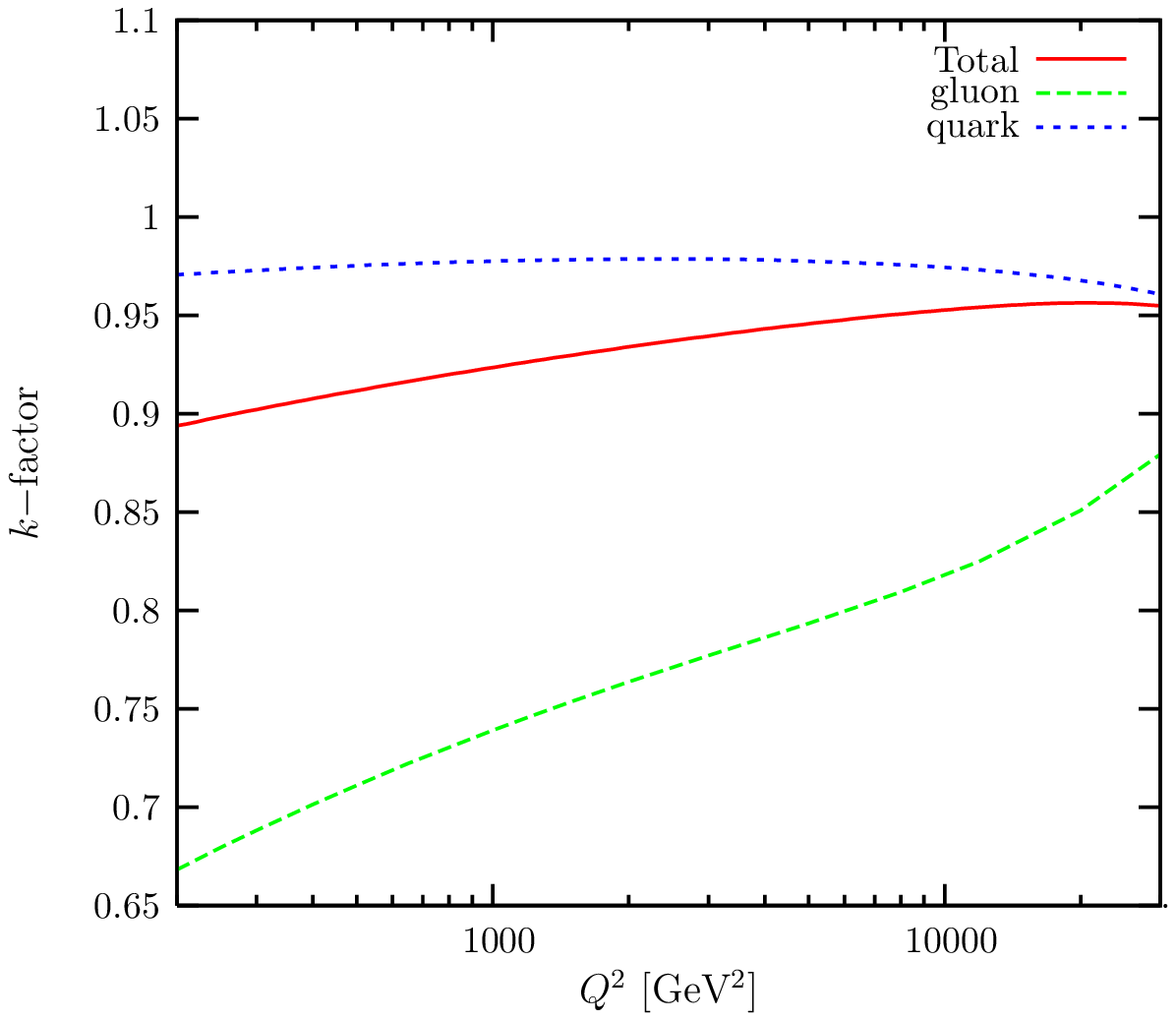}
\caption{Relevant $k-$factors for the contributions from initial state quark and gluons.}
\label{kfactor}
\end{minipage}
\end{figure}

 We first show the effect of the NLO corrections to pure graviton exchange, splitting the cross section into contributions from intitial state quarks and gluons. The calculation is performed using the NLO PDF set MSTW2008NLO \cite{Martin:2009iq}, with the factorisation and renormalisation scales set to $Q$. In Fig.~\ref{diffcro} we show the LO and NLO results for the differential cross section $d\sigma/dQ^2$ for $M_S=1$~TeV and the HERA positron and proton energies. This result is proportional to $M_S^{-8}$, so one can appropriately rescale for other values of $M_S$.
In Fig.~\ref{decomp} we decompose the results to gluon and quark initial state contributions while in Fig.~\ref{kfactor} we show the relevant $k-$factors where $k$ is defined as the ratio of the NLO cross section calculated with NLO PDFs to the LO cross section calculated with LO PDFs. We notice that the NLO corrections reduce the total cross section by up to 10$\%$, with the reduction being more significant for the gluon processes. This is due to the fact that the $\mathcal{O}(\alpha_s)$ corrections are dominated in both cases by large and negative contributions proportional to $\delta(1-x)$. For both LO and NLO cross sections, the dominant contribution comes from quark scattering, as we are considering the cross section at high values of the momentum fraction $x$ where the gluon PDF suppresses the gluon contribution. 

\begin{figure}[h]
\begin{minipage}[b]{0.5\linewidth}
\centering
\includegraphics[trim=1.5cm 0 0 0,scale=0.6]{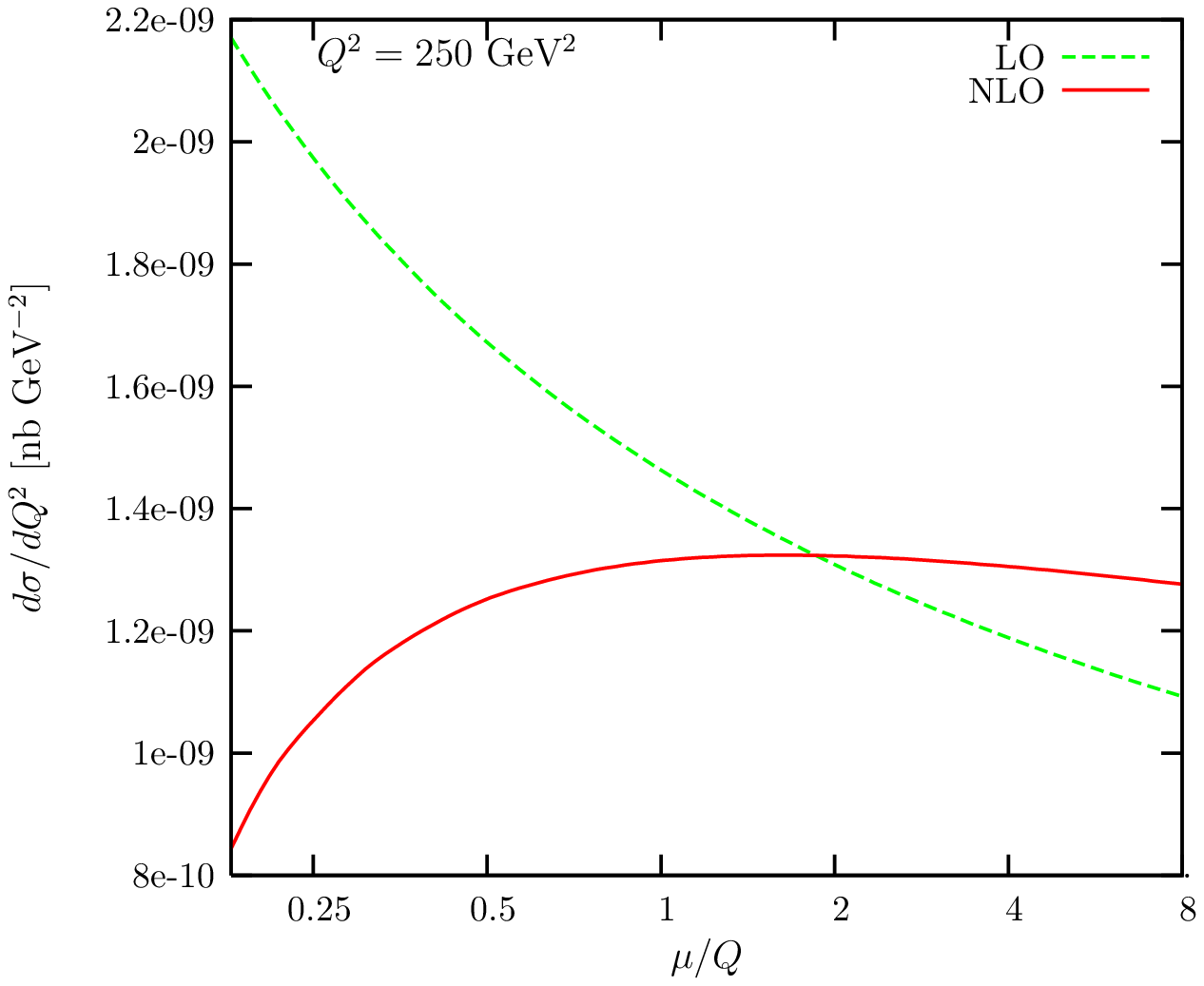}
\caption{Scale dependence of the differential cross section $d\sigma/dQ^2$ at $Q^2=250$~GeV$^2$.}
\label{scale1}
\end{minipage}
 \hspace{0.5cm}
\begin{minipage}[b]{0.5\linewidth}
\centering
\includegraphics[trim=1.5cm 0 0 0,scale=0.6]{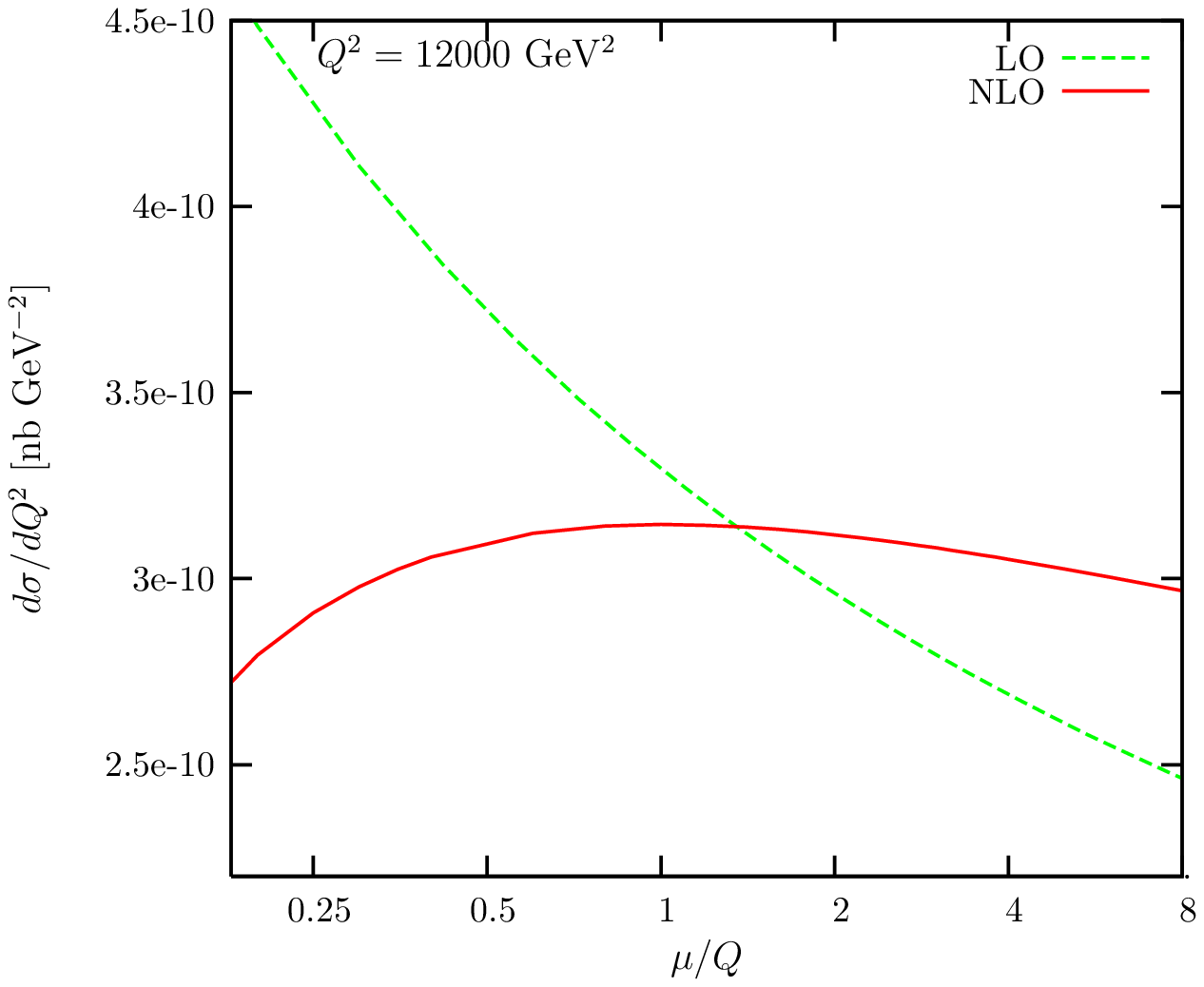}
\caption{Scale dependence of the differential cross section $d\sigma/dQ^2$ at $Q^2=12000$~GeV$^2$.}
\label{scale2}
\end{minipage}
\end{figure}

The advantage of the NLO calculation is that the result for the cross section is less dependent on the factorisation and renormalisation scales, and therefore the theoretical uncertainty is reduced. The dependence of the results on the scale choice is calculated by reinstating the appropriate $\text{log}(Q^2/\mu^2)$ factors in the coefficient functions. This can be seen in Figs.~\ref{scale1} and \ref{scale2} where we show the dependence of the LO and NLO cross section on the choice of factorisation and renormalisation scale which here are chosen to be equal and varied as a multiple of the momentum transfer $Q$. We consider two values of $Q^2$ to show the differential cross section for $M_S=1$~TeV, $Q^2=250$~GeV$^2$ and 12000~GeV$^2$. Similar results are obtained for all values of $Q^2$, and therefore the total cross section follows the same behaviour in terms of the scale dependence. 

\begin{figure}[h]
\centering
\includegraphics[trim=1.3cm 0 0 0,scale=0.65]{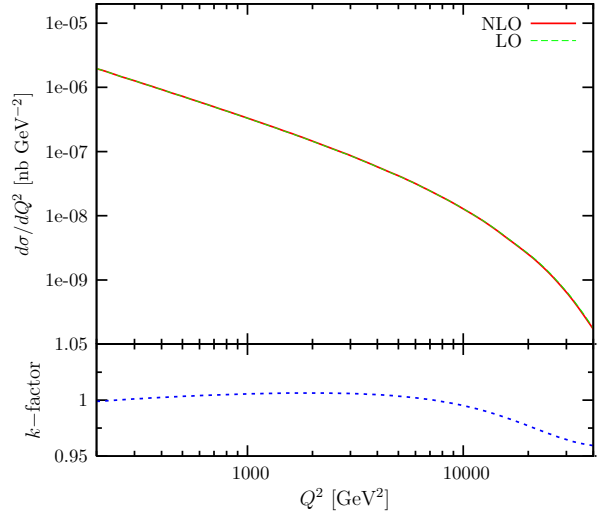}
\caption{Differential cross section from interference terms for $M_S=1$~TeV, and the corresponding $k-$factor.}
\label{inter}
\end{figure}

The effect of the NLO QCD corrections is also considered for the interference terms. The $k-$factor in this case is found to be very close to one as shown in Fig.~\ref{inter} for $M_S=1$~TeV. This is due to the fact that the pure signal contribution receives a negative contribution from NLO corrections while the SM prediction is increased by a similar relative amount when we consider order $\alpha_s$ corrections.

\begin{figure}[h]
\centering
\includegraphics[scale=0.7]{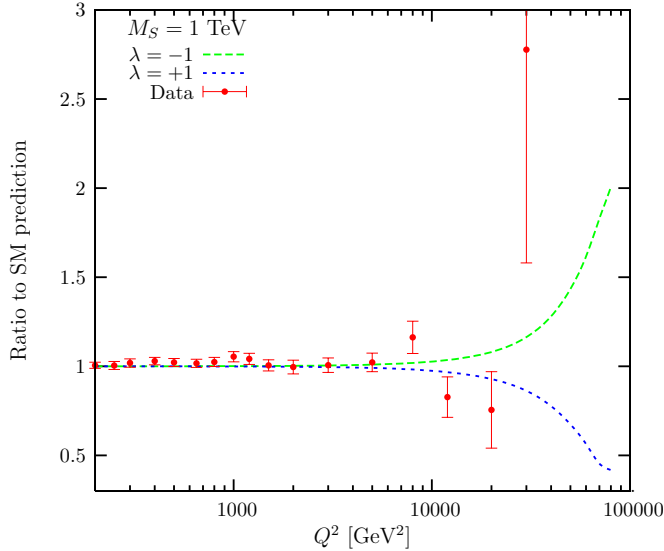}
\caption{Ratio of differential cross sections for $M_S=1$~TeV.}
\label{smratio}
\end{figure}

In Fig.~\ref{smratio} we show the ratio of the total differential cross section $d\sigma/dQ^2$ for the NLO new physics prediction to the NLO SM prediction. The data shown here are taken from \cite{Adloff:2003uh}. The experimental errors on the data points are the total errors given in column~7 of Table~6 in  \cite{Adloff:2003uh}. We note that for $M_S$=1~TeV we would only observe the interference effects which have opposite sign for $\lambda=\pm 1$. At sufficiently high $Q^2$ the pure graviton contribution takes over, giving a positive deviation from the standard model prediction. However the cross section falls rapidly with $Q^2$ and the values fall beyond measurable event rates.

\begin{figure}[h]
\begin{minipage}[b]{0.5\linewidth}
\centering
\includegraphics[trim=1.3cm 0 0 0,scale=0.65]{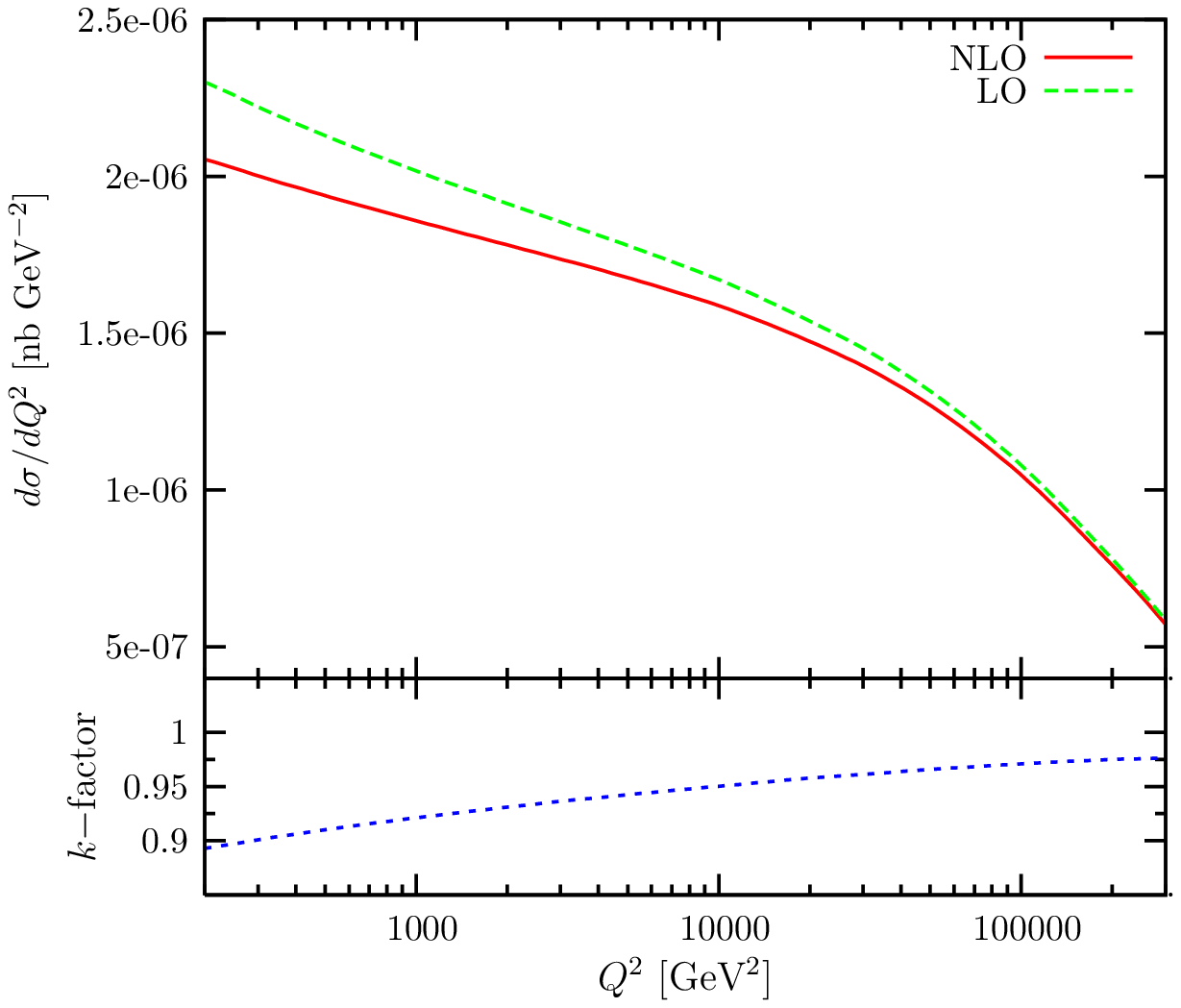}
\caption{Differential cross section for the LHeC at LO and NLO for $M_S=1$~TeV and the relevant $k-$factor.}
\label{lhecr}
\end{minipage}
\hspace{0.5cm}
\begin{minipage}[b]{0.5\linewidth}
\centering
\includegraphics[trim=1.7cm 0 0 0,scale=0.65]{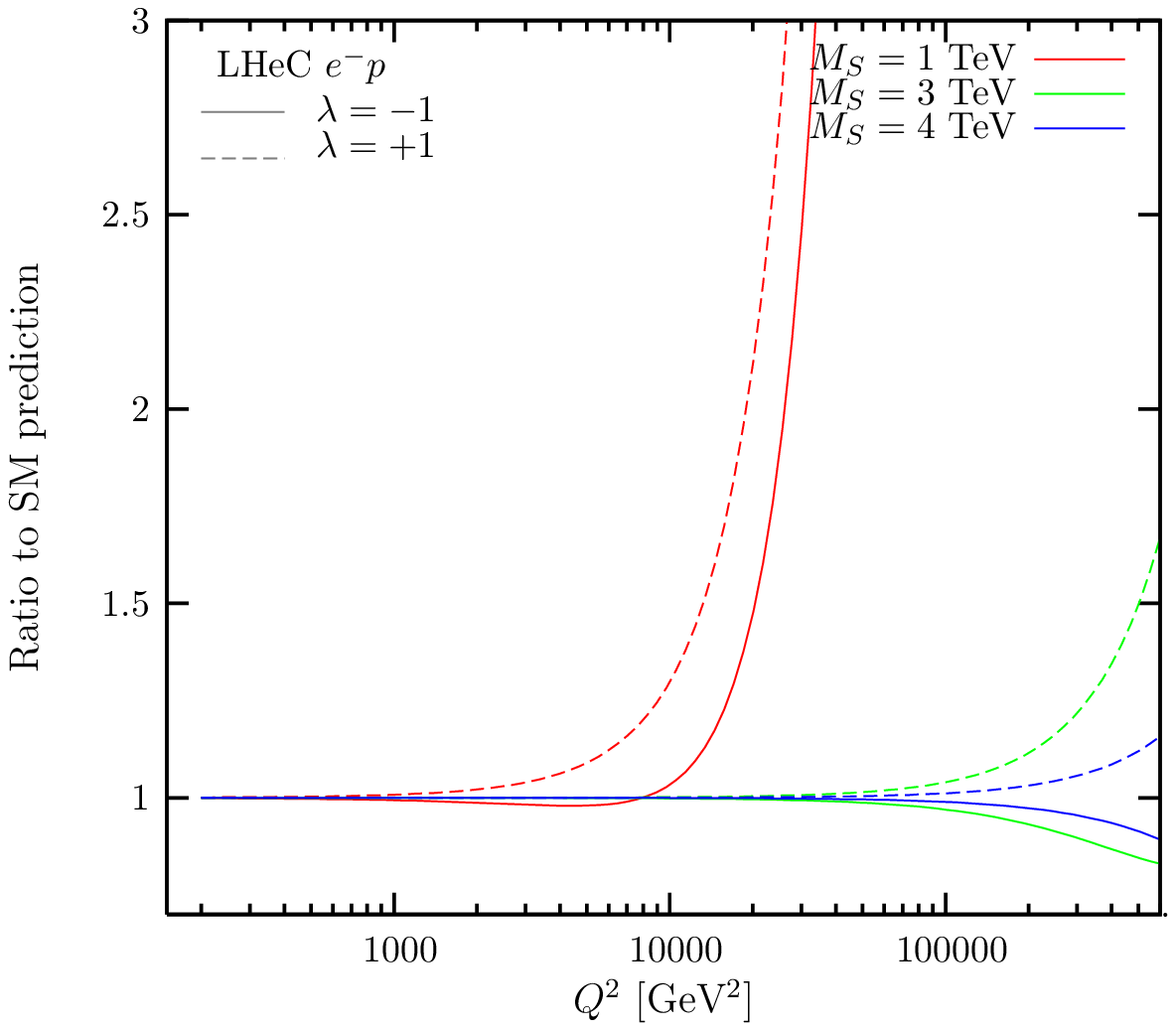}
\caption{Ratio of the NLO new physics and SM differential cross sections at LHeC energies for $M_S=1,3,4$ TeV. }
\label{lhec}
\end{minipage}
\end{figure}

Given the recent LHC results from dijet measurements~\cite{Collaboration:2010eza,Khachatryan:2010te} and the subsequent analysis of data which excludes fundamental scales up to $\sim3$~TeV \cite{Franceschini:2011wr}, the graviton contribution to DIS at HERA energies could not be detected. While the reach of HERA experiments is limited by the low centre-of-mass energy, the situation is more promising for the proposed LHeC. In Fig.~\ref{lhecr} we show the cross section for collisions of 7~TeV protons and 140~GeV electrons, keeping the same maximum $y$ value as at HERA for comparison with Fig.~\ref{diffcro} and the relevant $k-$factor. We note that the $k-$factor does not vary significantly between the two different centre-of-mass energies. In Fig.~\ref{lhec}, we show the deviation from the SM prediction for $M_S=1$~TeV for comparison with the prediction in Fig.~\ref{smratio} for HERA and to show the extended reach of the experiment also for $M_S=3$ and 4 TeV. We use different colours to plot the results for different scales and solid and dotted lines to show the effect of changing the sign of the interference terms. We notice that even for a fundamental scale as large as $M_S=4$ TeV, we can get a measurable deviation from the SM prediction, with the cross-section values now much higher leading to measurable event rates. Note that as the sign of the $\gamma G$ interference term is different the picture is slightly different from Fig.~\ref{smratio}, as the first deviation (lowest $Q^2$) from the SM prediction comes from the interference term.
\section{Conclusions}
In this paper we have studied the graviton contribution to DIS, defining appropriate structure functions and obtaining analytic results for the order $\alpha_s$ differential cross section. We calculate the relevant $k-$factors and find that NLO QCD corrections decrease the cross section by up to 10$\%$ with quark scattering contributing most of the cross section in the relevant region of high $Q^2$. We also show that, as expected, the NLO results are less dependent on the choice of factorisation and renormalisation scales and therefore decrease the theoretical uncertainty. We conclude that NLO corrections would only slightly modify the limits set on the fundamental scale, compared with LO calculations, as the limits depend mostly on the interference of the SM and graviton contributions. Given currents constraints set recently by LHC results, the graviton contribution to DIS is not within the HERA reach. However, in the context of the proposed LHeC experiment, given the significantly higher center-of-mass energy the graviton contribution to DIS could be observed for scales as high as 4~TeV. We have calculated the cross section and the expected deviation from the SM prediction at NLO with the calculated $k-$factor showing a similar behaviour as for the lower HERA energies. 
\acknowledgments{E.V. acknowledges financial support from the UK Science and Technology Facilities Council.}

\end{document}